# A New Approach to Radiocarbon Summarisation:

# Rigorous Identification of Variations/Changepoints in the Occurrence Rate of Radiocarbon Samples using a Poisson Process

**Authors:** Timothy J Heaton[1]\*, Sara Al-assam[1], Edouard Bard[2]

**Affiliations:**

[1]Department of Statistics, School of Mathematics, University of Leeds; Leeds, LS2 9JT, UK
\*Corresponding author. Email: t.heaton@leeds.ac.uk

[2]CEREGE, Aix-Marseille University, CNRS, IRD, INRAE, Collège de France; Technopole de l'Arbois BP 80, 13545 Aix en Provence Cedex 4, France.

**Abstract:** A commonly-used paradigm to estimate changes in the frequency of past events or the size of populations is to consider the occurrence rate of archaeological/environmental samples found at a site over time. The reliability of such a "*dates-as-data*" approach is highly dependent upon how the occurrence rates are estimated from the underlying samples, particularly when calendar age information for the samples is obtained from radiocarbon ($^{14}$C). The most frequently used "*$^{14}$C-dates-as-data*" approach of creating Summed Probability Distributions (SPDs) is not statistically valid, or coherent, and can provide highly misleading inference. Here, we provide an alternative method with a rigorous statistical underpinning that also provides valuable additional information on potential changepoints in the rate of events. Furthermore, unlike current SPD alternatives, our summarisation approach does not restrict users to pre-specified, rigid, summary formats (e.g., exponential or logistic growth) but instead flexibly adapts to the dates themselves. Our methodology ensures more reliable "*$^{14}$C-dates-as-data*" analyses, allowing us to better assess and identify potential signals present. We model the occurrence of events, each assumed to leave a radiocarbon sample in the archaeological/environmental record, as an inhomogeneous Poisson process. The varying rate of samples over time is then estimated within a fully-Bayesian framework using reversible-jump Markov Chain Monte Carlo (RJ-MCMC). Given a set of radiocarbon samples, we reconstruct how their occurrence rate varies over calendar time and identify if that rate contains statistically-significant changes, i.e., specific times at which the rate of events abruptly changes. We illustrate our method with both a simulation study and a practical example concerning late-Pleistocene megafaunal population changes in Alaska and Yukon.

**Highlights:**
- Summed probability distributions (SPDs) do not provide a valid, or statistically coherent, approach to summarize sets of $^{14}$C determinations.
- We introduce a statistically-rigorous and robust, fully-Bayesian, alternative that ensures more reliable *$^{14}$C-dates-as-data* analysis.
- Information on the varying occurrence rate of archaeological/environmental $^{14}$C samples over calendar time is provided
- The calendar timings of any substantial changes in the sample occurrence rate are identified and estimated.

- Code, vignettes and a user guide are available in the *carbondate* R library on CRAN and on Github at https://tjheaton.github.io/carbondate/

# 1 Introduction

Computational analyses of large, collated sets of radiocarbon dates have the potential to provide key inference on our past, on rates of change, and on population dynamics. It is however essential that the methods that underpin these "big-data" analyses are rigorous, robust, and reliable. This reliability is particularly critical in the case of radiocarbon, the most frequently-used method to date the last 55,000 years, as fluctuations in past atmospheric $^{14}$C levels (Heaton et al., 2021) mean that all radiocarbon determinations must be calibrated against an appropriate calibration curve to place them on the calendar age scale (Heaton et al., 2020; Hogg et al., 2020; Reimer et al., 2020). This calibration, to transform from radiocarbon determination to calendar age, introduces considerable challenges for any subsequent inference as, due to wiggles and plateaus in the calibration curves, the resultant calendar age estimates can have complex uncertainties and may even be multi-modal (Bronk Ramsey, 2009; Heaton et al., 2024; Reimer et al., 2025).

In this paper, we consider the challenge of rigorously summarising *$^{14}$C-dates-as-data* (Rick, 1987) to replace the statistically invalid methodology of summed probability distributions (SPDs). Importantly, unlike other approaches that have been developed to try and tackle this problem (e.g., Carleton, 2021; Crema & Shoda, 2021; DiNapoli et al., 2021; Price et al., 2021; Timpson et al., 2020) we seek a method that does not require a user to specify *a priori* a particular fixed form for that summary (e.g., exponential/logistic growth in the number of $^{14}$C samples over calendar time). In many circumstances, limiting the summary to such a rigid parametric structure will be inappropriate and result in the loss of key information. Instead, we aim to adapt our summary dependent upon the data themselves, using a non-parametric approach to create a final summarisation that is more flexible and more widely applicable. As such, our approach seeks to provide a true replacement for SPDs.

Specifically, suppose that we have collated a set of $n$ related archaeological/environmental samples, each of which is accompanied by a $^{14}$C determination (see Fig. 1). These samples might consist of charcoal from fires, human/animal bones, or other evidence of occupation found at a site over time. We aim to estimate, without making strong assumptions about its structure, the varying occurrence rate of these collated samples over calendar time; as well as analyse whether there are specific calendar ages at which the occurrence rate changes and, if so, when. *If* the creation of the collated samples is representative of some underlying process, the sample occurrence rate can provide a valuable proxy for activity level, population size, or environmental change. Time periods with a high sample occurrence rate may indicate an increased level of activity/population, while those periods with a low sample occurrence rate may suggest reduced activity/population. Variations and abrupt changes in the rate of observed samples might also suggest the influence of important external and environmental factors.

Most current *$^{14}$C-dates-as-data* summarisation is performed via summed probability distributions (SPDs): each $^{14}$C determination is first calibrated independently; and then the individual calendar age estimates are then summed. These SPDs are typically presented as though they provide summary estimates of the changing density of dates over time and then used as a proxy for changes in population size. However, it is well recognised that SPDs have considerable issues (Carleton & Groucutt, 2020; Contreras & Meadows, 2014; Crema, 2022; Crema & Bevan, 2021; Heaton, 2022; Michczyński & Michczynska, 2006). Indeed, in this paper we aim to show that SPDs cannot be relied upon to provide accurate estimates of temporal variations in the calendar age density of samples and should not therefore be used for archaeological/environmental inference. We propose, in contrast, an alternative that is not only fully rigorous but also provides valuable additional information on the timing of significant changes for use in later inference.

Our work is also of potential relevance to other geochronology communities that use SPDs, or their equivalents. This includes fission-track dating and OSL dating where it is argued that SPDs cannot properly recover two components of a mixture distribution (Galbraith, 1988, 1998, 2010). Additionally, those measuring the exposure duration of various rock surfaces (moraines, fault planes…) based on the *in-situ* production of cosmogenic nuclides ($^{10}$Be, $^{26}$Al, $^{36}$Cl and $^{14}$C) who currently use SPDs, or rather PDPs as they are known in this community (Schimmelpfennig et al., 2014; Vermeesch, 2012).

Summed probability distributions possess several critical flaws. Firstly, they are not statistically coherent. When creating an SPD, the calibration of each radiocarbon sample is performed entirely independently, and separately, both from all the other samples as well as the ensuing summarisation. This independent sample calibration is entirely in conflict with the concept that the samples are related to one another through some underlying process. When summarising, the fundamental assumption is that the samples arise from some shared calendar distribution. This should inform the calibration of the samples.

Secondly, the summary estimates that SPDs generate are overly variable, making inference extremely difficult, in particular when the calibration curves contain inversions, plateaus and substantial wiggles. It is straightforward, see Section 3, to show that they fail to reconstruct population size in simulations (Contreras & Meadows, 2014; Galbraith, 1988; Heaton, 2022). Attempts to address these issues, such as smearing out the calendar ages (Brown, 2015) do not resolve these problems but simply introduce other biases.

Finally, SPDs lack a method to rigorously quantify the uncertainty on their values. While bootstrap-based uncertainty estimates have been proposed to try and address this (Fernández-López de Pablo et al., 2019; Rick, 1987; Shennan et al., 2013; Timpson et al., 2014; Zahid et al., 2016) they do not satisfy the necessary requirements for consistency and hence are not valid or reliable as we also show in Section 3. In some instances, they may even make the inference worse by suggesting spurious precision or accuracy in the SPD. Due to these critical failings, we advise that SPDs should no longer be used by the archaeological and environmental science communities.

Bronk Ramsey (2017) has proposed two alternatives to SPDs that draw on ideas from frequentist kernel density estimation (KDE). Kernel density plots (*KDE_Plot*) simply convolve the SPD with a range of Gaussians, effectively just smearing out the original SPD. This does not address the fundamental issues of SPDs. Kernel density models (*KDE_Model*) offer an improvement in several areas. The KDE model approach iterates between (non-parametrically) creating interim estimates of the calendar age density of the collated $^{14}$C samples, and then recalibrating the samples using that interim estimate. This is repeated until convergence is achieved within an MCMC-type algorithm. However, the updating steps within this iterative algorithm remain somewhat inconsistent with a formal MCMC framework and are not underpinned by the necessary, and complete, Bayesian model. This makes it challenging to assess the robustness of the summaries generated by KDE models.

Fully formal, and rigorous, SPD-style $^{14}$C summarisation is instead available via a Dirichlet Process Mixture Model (DPMM) estimate (Heaton, 2022). This DPMM approach both summarises the joint distribution of the $^{14}$C samples over calendar time, and simultaneously estimates the number of calendar age clusters from which the samples arise, within a complete Bayesian framework. This has substantial implications for our ability to reliably estimate and reconstruct the underlying calendar age distribution of the population of samples, and to infer what this might mean in terms of past activity. In a DPMM, the set of radiocarbon samples are modelled as arising from an unknown mixture of such calendar age clusters. The number of clusters, along with their calendar period and spread, are adaptively estimated directly from the collated set of $^{14}$C observations within a fully-Bayesian MCMC framework. It is available in R using the *carbondate* library (and will shortly also be implementable in *OxCal*).

Here, we propose an alternative, but linked, approach for summarising calendar age information from multiple $^{14}$C samples. Given a set of $n$ samples, each with a radiocarbon determination, we aim to reconstruct the sample occurrence rate over time and variations within that rate. Our method retains all the statistical rigour of the DPMM summarisation approach (Heaton, 2022) but approaches the modelling of sample occurrences via an inhomogeneous Poisson process. Inference is performed within a fully-rigorous Bayesian framework using reversible-jump Markov Chain Monte Carlo (RJ-MCMC) (Green, 1995). The estimated sample occurrence rate provides information analogous to that intended by an SPD (or a DPMM) with the potential to act as a proxy for underlying population activity. In addition, the method identifies if there are statistically significant changepoints in the occurrence rate (i.e., specific times at which the rate of events abruptly changes). These times may indicate key changes in important external environmental factors. Such, highly-valuable, changepoint information is not available through traditional SPD-type summarisation.

Finally, we note that a range of parametric SPD alternatives have also been proposed. These require users to specify, in advance, a specific form/structure for the summary. While such models are useful if the data do indeed follow this structure, these methods are inherently much more restrictive in their field of application and provide less information on potential signals present. Furthermore, where the data do not follow the pre-specified structure, they will lead to inappropriate inference and, as we discuss in Section 3, the use of Monte Carlo-based SPD model tests (Shennan et al., 2013; Timpson et al., 2014) to infer deviations from the specified forms is not reliable. Timpson et al. (2020) propose a changepoint-style method where the summary consists of continuous piecewise linear sections. They also suggest a heuristic approach to determine the appropriate number of such sections. However, the final implementation requires a user to fix this number and uncertainty in its value is not incorporated into the eventual summary. Price et al. (2021) propose an approach where the summary consists of a finite mixture of Gaussian components, but with the user having to specify the number of components in advance. Our DPMM approach (Heaton, 2022) can be seen as a more flexible version of this approach – treating the number of Gaussian components as uncertain and variable during MCMC fitting. Additionally, Haslett & Parnell (2008) propose a Gaussian mixture summary in their BCHRON library although we would not recommend this summarisation approach as key aspects (specifically, the means and variances of the Gaussian components) are arbitrarily set during initialisation and not updated during fitting. A range of methods to fit specific growth models have also been proposed as SPD replacements (e.g., Carleton, 2021; Crema & Shoda, 2021; DiNapoli et al., 2021) which sit alongside more traditional phase modelling summaries (Bronk Ramsey, 2009). Again, these growth models require a user to specify in advance the particular form of summary they wish to fit (e.g., exponential or logistic growth). Determining an appropriate form will be a challenge for many applications, and in many circumstances no simple form will be appropriate. The use of non-parametric summarisation approaches, i.e., the Poisson process approach presented here and the DPMM approach of Heaton (2022), provide much more flexibility and the potential for more insightful inference.

Our paper is laid out as follows. In Section 2 we begin by discussing the broad challenges of any *dates-as-data* inference, discuss the need for representativeness of the samples to be summarised, and provide general user recommendations. In Section 3, we explain, in detail and with illustrative examples, why SPDs fail to provide reliable summarisation in even simple *$^{14}$C-dates-as-data* situations. Here, we discuss why this failure also has consequences for those using Monte Carlo-based SPD model tests to infer deviations from particular growth models. We then provide an intuitive explanation of our alternative Poisson process summarisation approach in Section 4 and briefly detail the MCMC fitting mechanism (with more detail provided in the Appendix). Section 5 shows the results of our Poisson process approach. We consider three simulated examples as well as several real-life datasets. Finally, in Section 6 we summarise our work and identify some areas for future development.

**Notation:** As standard in the radiocarbon literature, all ages in this proposal are reported relative to mid-1950 AD (= 0 BP, before present). The pre-calibration (observed) $^{14}$C ages/determinations, $X_i$, are given in units *$^{14}$C yr BP*. Calendar (or calibrated) ages, $\theta_i$, are denoted as *cal yr BP* (or sometimes as *cal AD/BC* if this is more appropriate for the calendar date interpretation). Quoted analytical uncertainties (denoted ±) on measurements are given at 1σ.

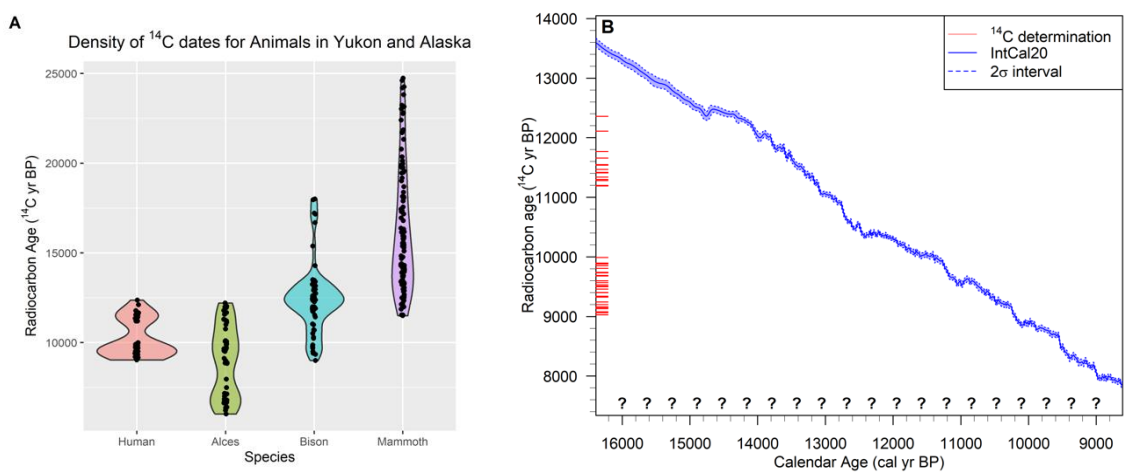

*Figure 1 Illustration of the $^{14}$C-dates-as-data challenge. Panel A: $^{14}$C determinations related to human occupation and three species of megafauna (Alces, bison, and mammoth) from Yukon and Alaska during the late Pleistocene (Guthrie, 2006). We wish to infer the underlying changing sizes of the various populations over calendar time, using the sample occurrence rates as a proxy. Estimation of the sample occurrence rates, and potential changes in their values, cannot be performed by directly analysing the frequency of the observed radiocarbon ages due to the need to calibrate these $^{14}$C determinations to place them on the calendar age scale. Events occurring at evenly-spaced calendar intervals may become highly uneven when considered in radiocarbon age as a consequence of wiggles, plateaus, and variations in the slope of the radiocarbon calibration curve. Panel B: $^{14}$C determinations for humans (red ticks) and the relevant calibration curve – blue, IntCal20 (Reimer et al., 2020). We must calibrate these determinations and estimate the underlying sample occurrence rate simultaneously.*

## 2 Recommendations for Analysing Big Datasets

We provide some general advice (and notes of caution) for applying any *dates-as-data* analysis. Big data analyses should not be used as black-boxes with the idea that the large amounts of data on which they are based mean that potential issues regarding the underlying sampling are less relevant and can be ignored – rather, the abundance of data make potential sampling issues ever more critical. Fundamentally, all users must consider and clearly define whether the data that they are summarising are representative of what they desire to study, or if there are potentially substantial sampling biases.

Several suggestions have been made in the literature about how users might automatically remove, or address, any sampling concerns yet still obtain useful proxies for activity or population size, such as binning/combining samples with similar $^{14}$C dates (Bevan et al., 2017; Crema & Bevan, 2021; Riris, 2018). While these may be appropriate in some situations, they do not work as automated approaches. Critically, the correct approach to take will be highly dependent upon the specific nature of the original sampling. Users of *$^{14}$C-dates-as-data* must therefore understand the methods of collection for the datasets they are seeking to summarise. There cannot be any automated procedure that cures all ills.

### 2.1 Representative Nature of Sampling

In order to use any *dates-as-data* analysis to obtain a proxy for population size or activity, it is critical that the set of samples to be summarised are representative of that population size/activity. Specifically, that the frequency, or rate of occurrence, over calendar time of the specific samples used for summarisation is proportional to that underlying size or activity. This is a strong requirement which must be considered by all users. It is relevant not only when using $^{14}$C samples but in any *dates-as-data* approach.

Archaeologists frequently have a desire to understand when a culture ended, began, or arrived in a location. This means they may be more likely to submit for $^{14}$C dating the youngest or oldest sample in a particular context rather than a random sample obtained from the site. They may also have a focus on a particular event and so submit more samples for $^{14}$C dating related to that. Equally, there are time periods (e.g., the Hallstatt plateau from c.a. 800-400 BC) where, due to plateaus in the calibration curve, $^{14}$C dating is perceived as being (or indeed is) unlikely to provide a sufficiently precise calendar date. Samples obtained from such periods may therefore not end up being submitted for $^{14}$C dating. Furthermore, databases of $^{14}$C dates may be dominated by projects with specific aims – such as to understand a specific site or a specific cultural phenomenon. The samples that are dated from these focussed projects may not be representative of the wider population. This is particularly an issue due to variations in project funding where larger projects will generate greater numbers of $^{14}$C dates that could swamp those from other projects. All these aspects may create substantial biases regarding which samples end up with $^{14}$C determinations in databases. These must be very carefully considered before implementing a *$^{14}$C-dates-as-data* analysis. Without doing so, we may spuriously end up providing inference on the nature of sampling (e.g., inferring that size/activity increased at the beginning/end of each cultural period) rather than the population itself.

### 2.2 Relationship between the $^{14}$C Samples and the Events of Interest

Any *dates-as-data* inference is also reliant upon the nature of the association between the calendar date of the samples and the archaeological/environmental events of interest (Griffiths et al., 2022; Waterbolk, 1971). Samples submitted for $^{14}$C dating may not be directly related, or indeed contemporaneous, with the event of interest. Instead, they may relate to a calendar time before, or after, the actual event of interest. For example, charcoal from a hearth may not correspond to the date of hearth creation, hafts from axes may not correspond to the date of manufacture, seeds from a grain store may not correspond to date of construction. Equally samples may only be loosely related to the events of interest and instead be residual or obtrusive. The relationship between the samples summarised and the events of interest remains of paramount importance.

## 2.3 Taphonomic Loss

In addition to the above sampling biases, we must also consider the potential for differential taphonomic loss whereby certain materials/cultures are less likely to be preserved or submitted for $^{14}$C dating (Bluhm & Surovell, 2019; Contreras & Codding, 2024; Ward & Larcombe, 2021). Older material may be more likely to be lost due to decomposition, transport or environmental conditions that affect how samples are preserved. Furthermore, certain materials are less likely to be submitted for $^{14}$C dating, while other materials may not be $^{14}$C-dateable at all. Again, this can lead to sampling biases in the specific set of $^{14}$C dates to be summarised where they do not represent the underlying population activity.

## 2.4 Failure of Automated Approaches to Address Sampling Biases

The most frequently-applied automated approach used to address sampling concerns in *$^{14}$C-dates-as-data* studies is to combine radiocarbon dates into bins before analysis (Bevan et al., 2017; Crema & Bevan, 2021; Riris, 2018). For example, replacing all samples at a site that sit within a 100 $^{14}$C yr interval with a single sample representing their mean. There are certainly instances when grouping together samples, and representing them with a single combined $^{14}$C date is appropriate. Indeed, there may be instances where this is required for useful inference (e.g., if some events generate multiple samples, while other do not). However, determining the necessary and appropriate process to identify such grouping must be done with great care and an awareness that doing so may also alter what the ensuing summary represents. Approaches that combine/bin samples based solely on their $^{14}$C dates are however problematic, even if that $^{14}$C-binning selection is done on a site-specific basis.

For example, it is likely essential to combine a group of $^{14}$C samples if one has external information that they all relate to a single event. Additionally, it may be sensible to group together samples that are known to relate to a single short-term occupation of a specific site by a group. Then the events of interest would become site occupation, and subsequent inference would relate to the occurrence rate of such occupations. Alternatively, one might propose it is reasonable to represent all the samples at a particular site by just the earliest sample – in which case, the event of interest (and subsequent inference) would become new site colonization.

However, binning/combining together samples based solely on the information that they have similar $^{14}$C dates, even if done site-by-site, will significantly bias all later inference. Equi-spaced $^{14}$C bin intervals do not equate to equal-length calendar intervals. By automatically binning together samples with similar $^{14}$C dates, the inherent effect will be to create estimates that have a lower event occurrence rate during flat/plateau periods of the $^{14}$C calibration curve (as the single binned sample will be spread over a long calendar time period) and a greater event occurrence rate during high gradient periods of the $^{14}$C calibration curve (as the single binned sample will occur in a narrow time interval).

Ultimately, to make reliable decisions on which samples to combine, and which to leave distinct, it is essential for the user to understand the nature of the samples themselves. This is unlikely to be automatable.

## 2.5 Potential non-identifiability of $^{14}$C calibration

Finally, the nature of $^{14}$C calibration necessarily means that calendar ages of some samples are not identifiable: when the curve exhibits plateaus or inversions. As illustrated in Fig. 3, a set of samples with $^{14}$C determinations around 2140 $^{14}$C yr BP could come from a short period of activity around 2300 cal yr BP, any interval of time between 2200 and 2000 cal yr BP, or a combination of these intervals. Due to the nature of the calibration process, we cannot distinguish which of these multiple options for the calendar timing of the activity is correct based the radiocarbon determinations alone.

Whenever performing *$^{14}$C-dates-as-data* we therefore strongly advise all users plot the $^{14}$C dates, and summary, alongside the calibration curve as shown in all figures within this paper. This will allow them to observe whether the presence of multiple peaks in the posterior mean occurrence rate are distinct activity periods or a reflection of inversions in the curve (resulting in potential non-identifiability of the activity period). In such instances users should be careful with interpretation and, if using an MCMC approach such as our Poisson process or DPMM, consider the individual posterior rate realisations (which should encompass a range of distinct fitting options) in addition to the summarised posterior mean.

We note that, in our Poisson process model, when presented with non-identifiability, which option will be preferred will depend upon the prior specified on the number and spacing of the changepoints. The default choice of prior will prefer the most parsimonious model (with fewer changepoints in the occurrence rate and without multiple, closely spaced, rate changes) to avoid overfitting but this can be altered by the user if external and independent information is available.

# 3 The Failure of Summed Probability Distributions (SPDs)

Given a set of $n$ $^{14}$C samples, $X_1, \ldots, X_n$, the most frequently-used approach for summarising their combined calendar information relies upon summed probability distributions (SPDs). We suppose that sample $i$, with $^{14}$C determination $X_i$, has unknown calendar age $\theta_i$. When summarising, we assume each $\theta_i$ arises from a shared common density $f(\theta)$ that we wish to estimate as our activity proxy.

The SPD approach does not however provide a coherent or reliable estimate of this shared $f(\theta)$. Below, we demonstrate why SPDs should consequently not be used for inference – illustrating their failure on simple examples and explaining the reasons behind that failure.

## 3.1 Creating an SPD

To create an SPD, one performs two distinct steps:

- Calibrate (independently) each determination $X_i$, to obtain an estimate $\hat{f}_i(\theta_i)$ for its unknown calendar age $\theta_i$.
- Create SPD $\hat{f}(\theta)$ by summing/averaging all these independent estimates:

$$\hat{f}(\theta) = \frac{1}{n} \sum_{k=1}^{n} \hat{f}_i(\theta_i)$$

While simple to construct, the resultant SPD $\hat{f}(\theta)$ is not a statistically-valid estimate of the underlying $f(\theta)$ we wish to reconstruct. The summaries they generate contain several fundamental flaws.

## 3.2 SPDs are overly-variable and over-spread

It is well-recognised that SPDs are overly-wiggly (Michczyński & Michczynska, 2006) introducing unwanted artefacts and peaks into the summary that reflect features of the calibration curve rather than the underlying calendar ages of the samples. They also tend to be overspread, making it look like activity begins earlier and ends later than it in fact does. This over-variability and over-spreading make SPDs extremely difficult to reliably interpret and understand – is a particular feature/peak in the SPD genuine or is it simply an artefact (Contreras & Meadows, 2014; Galbraith, 1988; Heaton, 2022)?

This is illustrated in Fig. 2A. Here, we have simulated 50 calendar ages $\theta_1, \ldots, \theta_{50}$ from the calendar age distribution shown in red (a mixture of two, smooth, normal phases centred at 5000 and 3500 cal yr BP respectively). We then simulate $^{14}$C determinations $X_1, \ldots, X_{50}$ corresponding to these simulated calendar ages. Given only the 50 $^{14}$C determinations, we aim to reconstruct the true (red) shared calendar age/activity distribution. While the SPD (shown in grey) is able to provide some general features of this underlying true distribution, it also contains many further artefacts and peaks of a similar size. In particular, spurious peaks are created that could be falsely interpreted as implying much greater variability in the underlying activity/population than that actually present in the smooth (red) truth. Additionally, the SPD values in the tails of both the phase centred on 5000 cal yr BP and (to a lesser extent) the phase centred on 3500 cal yr BP extend are overspread. These SPD values are substantially greater than the corresponding normal red tail values and extend further, giving an incorrect impression that the activity in each phase both began earlier and ended later than it did.

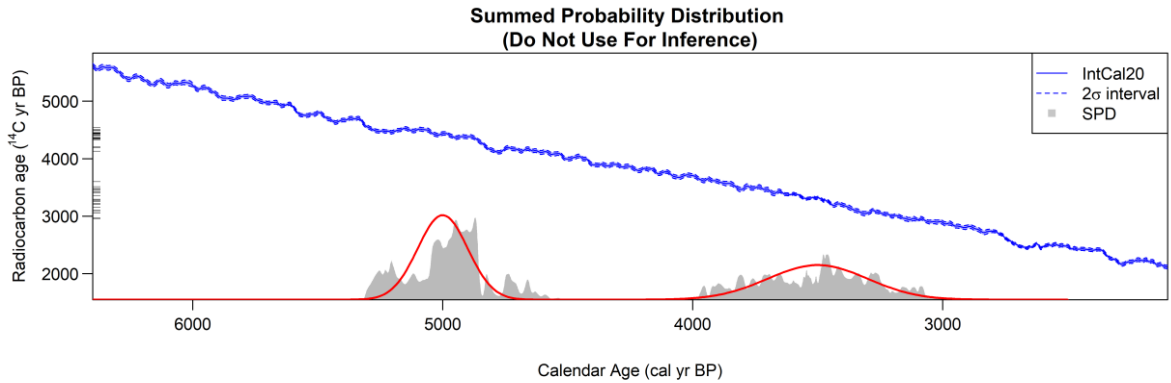

*Figure 2 Creating an SPD (shown as grey distribution) based on 50 simulated $^{14}$C determinations (black ticks on the radiocarbon age axis) known to be from samples with calendar ages drawn from a mixture of two smooth normal distributions/phases (red distribution). While the SPD recreates some broad features of the true (red) calendar distribution it is overly variable, overly spread, and introduces multiple further features, peaks and artefacts that are not present in the true (red) calendar age distribution.*

### 3.3    SPDs fail to fit any sort of model in the calendar age domain

The total lack of modelling in the calendar age domain is a substantial failing in SPDs. This is inherent to their creation – with complete independence between the calibration of each of the individual $X_i$ radiocarbon determinations and the simple averaging of the resultant calendar age distributions. This flawed independence, and lack of modelling of the data creation process, is partly what lies behind the over-variability and over-spreading discussed above. It can also lead to significant further issues and highly spurious interpretations.

In Fig. 3, we provide a simple illustration of this flaw with a trivial example where we create an SPD from a single sample (with a $^{14}$C determination of 2141 ± 30 $^{14}$C yr BP). With a single sample, we evidently have no information as to whether there are multiple distinct periods of population activity, separated by non-active periods. Instead, we are simply unsure of precisely when that single sample arose. However, when one creates an SPD from this sample, it does generate multiple peaks. Under a standard SPD paradigm, a user would be directed to interpret these as indicating multiple distinct periods of activity. While this is a somewhat pathological example, being based upon a single sample, the same flaw holds for SPDs based on multiple samples. Fundamentally, the calendar ages of a set of samples with $^{14}$C determinations around 2141 $^{14}$C yr BP are not identifiable. They could arise from a period of activity around 2280 cal yr BP or 2100 cal yr BP, or a combination of both.

To be reliable, any *$^{14}$C-dates-as-data* summarisation method must incorporate some modelling of the underpinning process by which the collected samples are created in the calendar age domain. Such modelling can however remain non-parametric (or very loosely parametric) so that when estimating the process, we remain informed by the observed data itself as to its form.

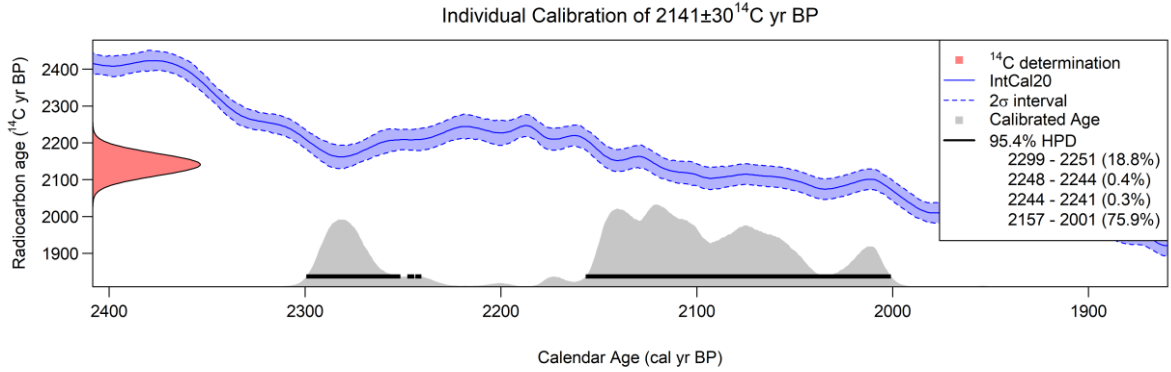

*Figure 3 An SPD based on a single $^{14}C$ determination indicating multiple SPD peaks. With a single sample, we have no information regarding the actual number of periods of activity – we are simply unsure as to the specific sample's calendar age. Interpreting the multiple peaks in the SPD as indicating disjoint periods of activity is flawed. While a trivial example, it aims to show the dangers of over-interpreting SPD peaks without consideration of the calibration curve, and the impact of a lack of modelling within SPDs.*

### 3.4 Failure of SPD bootstrap confidence intervals

The above failings would be somewhat ameliorated if it was possible to provide confidence intervals for SPDs that accurately represented the uncertainty on the underpinning summary. Attempts to do so have been made using bootstrapping (Fernández-López de Pablo et al., 2019; Shennan et al., 2013; Timpson et al., 2014; Zahid et al., 2016). However, these bootstrap methods are also not valid and can fail catastrophically, potentially even worsening the inference as they provide spurious ideas of precision and accuracy.

To demonstrate, in Fig. 4 we consider a simulated example where the underlying population activity occurs only in the 50-year interval from 2100-2050 cal yr BP (red shaded calendar age distribution, Fig 4A). We have generated 50 calendar ages $\theta_1, \ldots, \theta_{50}$ uniformly at random from this calendar interval. From these, we simulated corresponding $^{14}C$ determinations $X_1, \ldots, X_{50}$ (shown as ticks on the radiocarbon age axis) based upon the value of the IntCal20 calibration curve. We can immediately see from Fig. 4B that the SPD (shown in light shaded grey) generated from these 50 $^{14}C$ determinations $X_1, \ldots, X_{50}$ fails to reconstruct the underlying (light red) calendar age distribution from which the samples actually came.

This reconstruction infidelity in the SPD could potentially be acceptable if we were informed as to the SPD's uncertainty and were able to obtain confidence/probability intervals for the SPD that incorporated the true calendar age distribution. However, the suggested solution of bootstrapping the SPD (Fernández-López de Pablo et al., 2019; Rick, 1987; Shennan et al., 2013; Timpson et al., 2014; Zahid et al., 2016) does not provide such intervals. We cannot therefore be confident as to whether any SPD has provided an accurate estimate for the true calendar times that a group of $^{14}C$ samples were produced, or it has not. When bootstrapping an SPD based on $n$ radiocarbon samples, the process undertaken consists of first creating an initial SPD $\hat{f}(\theta)$ from the observed data (grey-shaded distribution, Fig. 4B). Then, to create a single bootstrap-resampled SPD estimate $\hat{f}^{boot,j}(\theta)$ we

1. Sample, from this initial SPD $\hat{f}(\theta)$, a new set of calendar ages $\theta_1^{boot,j}, \ldots, \theta_n^{boot,j}$. Such a potential set, for our example where $n=50$, are shown as purple ticks on the calendar age axis in Fig. 4B.

2. Sample a corresponding set of simulated $^{14}C$ determinations $X_1^{boot,j}, \ldots, X_n^{boot,j}$ based on these calendar ages, and the calibration curve (ticks on radiocarbon age axis, Fig. 4C)

3. Calculate a bootstrap-resampled SPD $\hat{f}^{boot,j}(\theta)$ from $X_1^{boot,j}, \ldots, X_n^{boot,j}$ (light purple distribution, Fig. 4C).

This process of resampling new sets of calendar ages from the initial SPD, $\hat{f}(\theta)$, followed by simulation of corresponding radiocarbon ages that are then re-summarised, is repeated many times to create a large number of bootstrap-resampled SPDs $\hat{f}^{boot,j}(\theta)$.

It has been proposed in the literature (Fernández-López de Pablo et al., 2019; Shennan et al., 2013; Timpson et al., 2014; Zahid et al., 2016) that this set of bootstrap-resampled SPD can then be used to provide relevant uncertainty intervals on the original SPD. For example, to estimate the 95% confidence interval on the SPD for a particular calendar year, it has been suggested we can simply choose the corresponding (2.5% and 97.5%) quantiles of these bootstrapped-resampled $\hat{f}^{boot,j}(\theta)$. However, Fig. 4D shows that this fails catastrophically. Not only does the bootstrap-based 95% interval (dotted line) fail to incorporate the underlying distribution that we originally aimed to reconstruct (the red uniform), but it also fails to even include the initial SPD based on our original data. Instead, the bootstrap-based intervals introduce even more spurious variation and wiggles into the summarisation, are even more overspread, and are even more strongly influenced by the nature of the calibration curve.

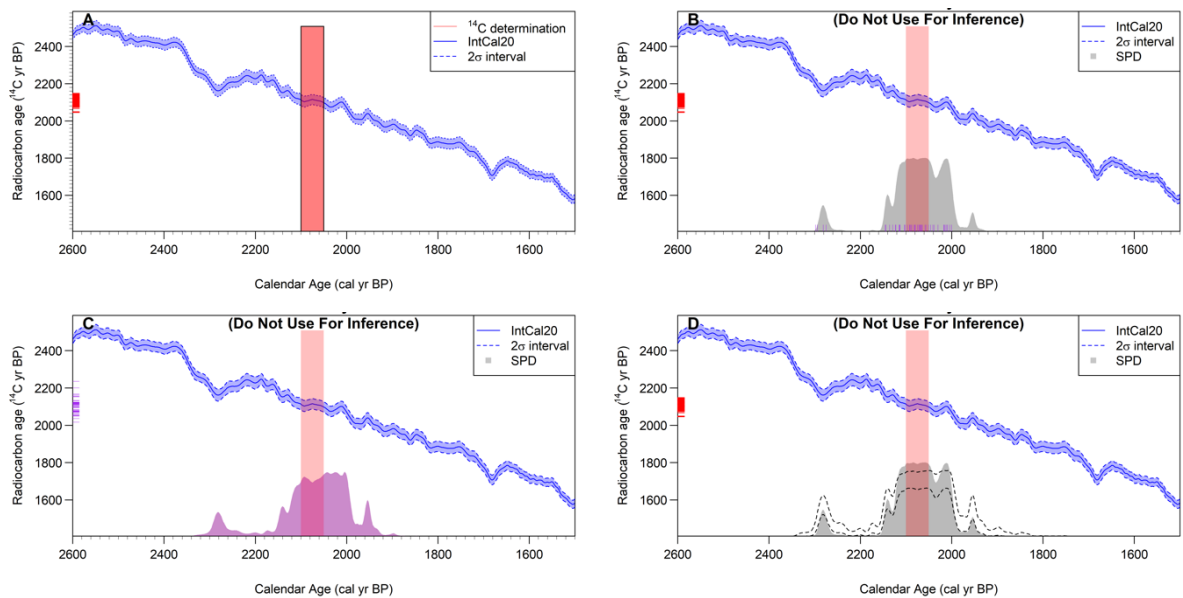

*Figure 4 Failure of SPD bootstrapping techniques. Panel A: Sampling $^{14}$C determinations (red ticks) for 50 simulated samples with calendar ages drawn from a known, uniform-phase, distribution (shaded red distribution) We aim to reconstruct this uniform-phase using $^{14}$C-dates-as-data. Panel B: The SPD (grey distribution) fails to reconstruct the underlying, true, distribution. For bootstrapping, we draw a set of 50 calendar ages from the initial SPD (purple ticks). Panel C: We simulate $^{14}$C determinations for these 50 resampled calendar ages (black ticks) and re-fit an SPD to them to obtain a bootstrap-resampled SPD estimate (light purple distribution). We repeat this process (of resampling calendar ages from the initial SPD in panel B, simulating corresponding $^{14}$C determinations, and SPD refitting) many times to obtain a large set of bootstrapped SPDs. It has been suggested these bootstrapped SPD can be used to provide SPD confidence intervals. Panel D: The initial SPD (grey distribution) with supposed 95% confidence intervals obtained from the bootstrapped SPDs (dashed lines). The intervals fail completely – not only do they fail to incorporate the underlying calendar age distribution (red), but they also fail to even include the original SPD.*

## 3.5 Why does bootstrapping fail for SPDs?

Bootstrapping is, for many statistical problems, a reliable way to understand the uncertainty of an estimator (Efron & Tibshirani, 1993). It does not however work for all problems. The reasons as to why bootstrapping fails for SPDs are multiple and complex. Fundamentally, success for a bootstrap-based SPD approach is reliant on the difference between the true $f(\theta)$ and the initial SPD $\hat{f}(\theta)$ being mimicked by the difference between $\hat{f}(\theta)$ and the bootstrapped SPDs $\hat{f}^{boot,j}(\theta)$. Intuitively, if this equivalence holds, what we learn about $\hat{f}^{boot,j}(\theta)$ as an estimator for $\hat{f}(\theta)$, which can be understood by creating a lot of bootstrap resamples, can then tell us about the initial $\hat{f}(\theta)$ as an estimator for the target $f(\theta)$. Unfortunately, for SPDs, such an equivalence does not hold. This can be seen in Figs. 4B and 4C. The difference between the initial SPD (light grey distribution) and the truth (red distribution) in Fig. 4B is not matched by the difference between the bootstrapped SPD (purple) and the initial SPD in Fig. 4C.

## 3.6 Implications of SPD Failure for Monte Carlo-based SPD model testing

The failure of SPDs to accurately recreate the calendar age distributions from which samples are generated also has ramifications for the reliability of inference in Monte Carlo-based SPD model testing (Shennan et al., 2013; Timpson et al., 2014), specifically their use to identify specific deviations from the chosen, null, models.

The basic concept of these Monte-Carlo tests, that, if the samples are generated according to a null model/distribution one is testing, then their SPD should lie within the Monte-Carlo intervals of SPDs based on data simulated from the null model, is valid. However, the Monte-Carlo SPD model testing approach only reliably provides a measure of the overall quality of fit. The level of dissimilarity between the observed SPD and the null model's Monte-Carlo SPD intervals must be assessed as a whole, ideally with a p-value to determine significance, and cannot easily be interpreted at period specific levels. The only robust test conclusions being to reject, or not reject, the complete null model as plausible.

Use of Monte-Carlo SPD tests to understand precisely how models might differ is much more challenging as inference on the SPDs does not directly equate to inference on the underlying calendar age distribution. Crucially, one cannot reliably interpret specific deviations between an observed SPD and the null-model's Monte-Carlo SPD intervals. It is not the case that, if the observed SPD lies above/below the null-model's Monte Carlo SPD intervals at some points, this necessarily implies there are more/fewer observations in that particular time period than the null calendar model would support.

Our Fig. 4 example demonstrates this. In Fig 4B, we show the SPD (grey distribution) obtained by simulating $^{14}$C data under a 2100-2050 cal yr BP uniform phase null calendar age model (red distribution). If we repeat this simulation step, using this uniform phase null, we obtain Monte-Carlo SPD intervals that are centred around this grey SPD. Critically, these Monte-Carlo SPD intervals bear little direct relation to the null calendar age model, with the SPD intervals having substantial non-zero values from 2150-2000 cal yr BP as well as distinct peaks at 2300 and 1950 cal yr BP (as well as several complex fluctuations and artefacts). This absence of a direct relationship immediately illustrates that aiming to interpret deviations between a null model's Monte Carlo SPD intervals and the SPD of observed $^{14}$C data is not meaningful. In our example, under such a flawed paradigm, we would potentially make inferential statements that an observed $^{14}$C dataset had fewer samples than possible under our null around 2000 and/or 2300 cal yr BP. However, such inference would evidently be incorrect. The uniform phase null already has zero probability at these times.

Users of Monte Carlo SPD tests should also be aware that, if they select a null model to test by fitting to the initial SPD then they may also end up with over-confidence and aspects of circularity in their inference. The initial SPD may not accurately reflect the true data generation mechanism and hence other, equally likely, data generating mechanism may not be fairly considered.

## 4 A rigorous Poisson process approach to summarisation

Our approach to *14C-dates-as-data* summarisation takes a different approach to SPDs. Intuitively, we suppose that the population we wish to summarise is continuously undertaking activity of potentially varying intensity. At specific times $\theta_1, \theta_2, ...$, this activity will generate an *event of interest*. Each of these *events* is assumed to create a radiocarbon sample. We model the occurrence of these events as random and aim to investigate whether, and how, the rate at which they occur varies over calendar time using the $^{14}$C samples that they leave behind. In the case of the late-Pleistocene megafauna used in our illustrative example, these events might be the death of an animal that leaves behind a bone to be dated. Note that we still require that the set of $^{14}$C samples to be summarised is representative of the underlying *events of interest* as described in Section 2.

### 4.1 An inhomogeneous Poisson process

Suppose we have collated a set of $n$ environmental/archaeological radiocarbon samples, with determinations $X_1, ..., X_n$, that have been generated by these underlying events. Specifically, we assume that the random calendar times $\theta_1, \theta_2, ...$ of these events occur according to an inhomogeneous Poisson process (PP) with a variable occurrence rate $\lambda(\theta)$.

A PP is a commonly used statistical approach to model random events distributed over time. The rate $\lambda(\theta)$ of the PP determines the number of events expected to occur in any particular time interval. Periods when the rate $\lambda(\theta)$ is high will typically be expected to generate a greater number of events; while periods when $\lambda(\theta)$ is lower will be expected to generate fewer events. An illustration is provided in Fig. 5, with full details in the Appendix.

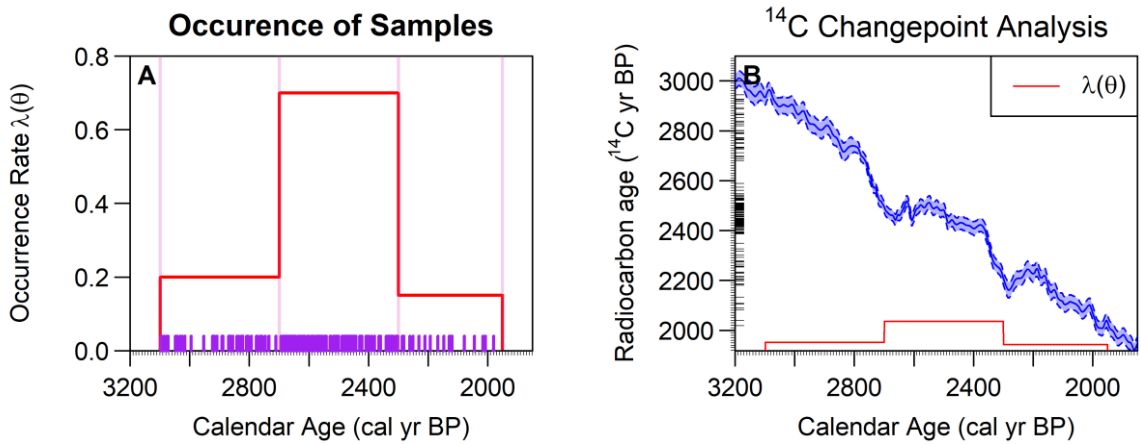

*Figure 5 – A Poisson process model for $^{14}$C-dates-as-data summarisation. Panel A: A PP with four changepoints in the occurrence rate (shown in red). Under a PP, events (purple) occur at random calendar times, but proportional to the underlying rate. Panel B: Each event creates a sample with a $^{14}$C determination (black ticks on radiocarbon age axis). We wish to reconstruct the underlying Poisson process rate, and its changes, given only the $^{14}$C determinations.*

### 4.2 Modelling the underlying sample occurrence rate

We model the rate $\lambda(\theta)$ as piecewise constant but with an unknown number of changepoints, specifically:

$$\lambda(\theta) = \begin{cases} h_i & \text{if } s_i \leq \theta < s_{i+1} \\ 0 & \text{otherwise}, \end{cases}$$

where $T_A < s_1 < s_2 < \cdots < s_k < T_B$ are the calendar ages at which there are changes in the occurrence rate, and $T_A$ and $T_B$ are bounding calendar ages outside of which it is known that no events occurred. These can be set as sufficiently distant from the data.

The number of changepoints $k$ in the underlying occurrence rate is considered unknown, as are the specific calendar times $s_1, s_2, \ldots, s_k$ of those changes and the actual value of the rate $h_1, \ldots, h_{k+1}$ in each interval. All these can vary within the model (see Appendix). While such a piecewise constant model for the sample occurrence rate might initially appear as a strong assumption, this model is highly flexible. Any function can be approximated by a suitable set of piecewise constant sections and, as we allow our model to vary the number, timings, and rates of these constant sections, we can reconstruct a very broad range of occurrence rates. It is not required that the true occurrence rate has a piecewise structure, as we show in our third simulation study of exponential growth and our late-Pleistocene megafaunal analysis.

### 4.3 Fitting the model to data using RJ-MCMC

We use Reversible Jump Markov Chain Monte Carlo (RJ-MCMC) through a Metropolis-within-Gibbs scheme that alternates between calibrating the samples $X_1, \ldots, X_n$ and updating the estimate of the underlying rate $\lambda(\theta)$:

**Step 1.** Update the calendar ages of the samples $\theta_1, \ldots, \theta_n$ given their $^{14}$C determinations $X_1, \ldots, X_n$ and the current estimate of the PP rate $\lambda(\theta)$. Given the rate $\lambda(\theta)$, the PP induces a prior on the calendar date of each sample that can be calculated exactly. Each $^{14}$C determination $X_1, \ldots, X_n$ is then calibrated using this shared calendar age prior.

**Step 2.** Update the PP rate $\lambda(\theta)$ given the current calendar ages of the samples $\theta_1, \ldots, \theta_n$. This step follows a standard RJ-MCMC procedure (Green, 1995) to estimate a PP rate given a set of events occurring at known times.

This fitting procedure is entirely automated within the *carbondate* library. See Appendix for full details. To assess MCMC convergence, we recommend that users run the sampler multiple times and visually compare the estimated posterior mean occurrence rates (and their probability intervals) for each run. If the chains have converged these should be similar. We also advise that users check that their results are robust to the prior on the number of changepoints (see Section 4.5).

### 4.4 Model Output

The MCMC generates a set of posterior realisations for the PP rate $\lambda(\theta)$ and the calendar age of each individual $^{14}$C determination. Most directly this allows us to estimate the sample occurrence rate $\lambda(\theta)$ at any calendar time. However, it also provides significant additional information including the number of distinct changes in the occurrence rate of the samples in the period studied, and the specific calendar ages at which such changes take place. Examples of how to access, use, and interpret such changepoint information are provided in Section 5. Furthermore, by drawing a large set of samples from the MCMC posterior, we are able to obtain rigorous posterior credible intervals for all these model components (the overall occurrence rate $\lambda(\theta)$, the number of changepoints, the location of the changepoints, …). We are also able to access and plot individual posterior realisations, as well as plot the posterior mean occurrence rate conditional on a specific number of changepoints should this be of interest - see Appendix.

### 4.5 Selecting model hyperparameters and priors

Within our PP model, we are required to place some priors on the values of parameters – in particular on the number of changepoints and their locations. Our default choice of prior means that our approach will prefer to fit parsimonious models where there are fewer changes in the sample occurrence rate (and changes do not occur in rapid succession) if such models are equally likely to explain the observed $^{14}$C determinations. This should help prevent over-fitting to random sampling variations and reduce the introduction of spurious rate changes. Users are however free to edit these choices if they have independent information – again, see Appendix for more information.

## 5  Results

To illustrate our method in practice we consider both simulated examples, where we simulate samples drawn from a known calendar age distribution that we then aim to recreate via *$^{14}$C-dates-as-data*, and a real-life example studying late-Pleistocene megafauna in the Yukon and Alaska (Guthrie, 2006). In all our examples, we run our MCMC for 100,000 iterations, discarding the first 50,000 as burn-in before thinning to every 10$^{th}$ sample. Convergence was evaluated by visually assessing the evolution of the posterior mean occurrence rate within the sampler, as well as rerunning with different initialisations.

### 5.1  Simulation Study

We consider three examples. In the first, we specify an underlying, shared, calendar age distribution $f(\theta)$ for our simulated samples. In the second and third, we simulate events (with corresponding samples) directly from a specific Poisson process. The generality of our modelling approach is such that any calendar age distribution $f(\theta)$ has an equivalent Poisson process but conditioned on $n$, the number of events/samples which occur. Hence, we can move easily and naturally between the two interpretations/paradigms.

In the first example, using our chosen calendar age distribution $f(\theta)$, we independently sample $n$ calendar ages $\theta_1, \ldots, \theta_n$ for a chosen value of $n$. In the second and third examples, where we sample events directly from a Poisson process, we will again generate calendar ages $\theta_1, \ldots, \theta_n$ but the number of events $n$ will itself be random. For each calendar age, $\theta_i$, we then create a corresponding $^{14}$C determination, $X_i$, using the IntCal20 calibration curve (Reimer et al., 2020) and a typical laboratory uncertainty of $\sigma_{obs} = 25$ $^{14}$C yrs. We apply our Poisson process summarisation approach to the set $X_1, \ldots, X_n$ and compare the estimate to the true $f(\theta)$. In each example, we also present the equivalent SPD.

Our first and second examples simulate from models that themselves have piecewise constant occurrence rates. However, our third example shows we are also able to successfully reconstruct a Poisson process corresponding to an exponential growth model (Crema & Shoda, 2021) demonstrating the approach's much broader applicability. For all our analyses, we chose the *carbondate* library default priors, in particular a prior mean of three internal changepoints.

#### 5.1.1  Example 1 – A single uniform phase (A Poisson process with two changepoints)

The data used here are the same as that used in Fig. 4 which illustrated the catastrophic failure of SPDs and their reported confidence intervals. We assume that the events (samples) must lie within the calendar period from [2350, 1850] cal yr BP. Within this period, we create 40 artificial samples with $^{14}$C determinations corresponding to the (narrower) calendar period from [2100, 2050] cal yr BP. Effectively, this is equivalent to a Poisson process with a rate:

$$\lambda(\theta) = \begin{cases} h_1 = 0 & if\ 2350 \geq \theta > 2100\ cal\ yr\ BP, \\ h_2 & if\ 2100 \geq \theta > 2050\ cal\ yr\ BP, \\ h_3 = 0 & if\ 2050 \geq \theta > 1850\ cal\ yr\ BP, \end{cases}$$

but conditional on there being 40 events occurring. The results are shown in Fig. 6. This example is particularly challenging as the calibration curve has multiple wiggles in this time period. These wiggles increased the seriousness of the failings when applying an SPD (Fig. 4). We can however see that the PP approach performs much better. The posterior for the PP rate $\lambda(\theta)$ does satisfactorily recreate the key aspects of the underlying model, and the 95% credible interval encapsulates the distribution from which the samples were actually created (Fig. 6A). We can also see (Fig. 6B) that the PP model estimates that the most likely number of changes in the sample occurrence rate is indeed two, although three internal changes is also supported. Additionally, Fig. 6C indicates that, conditional on two changepoints, the model estimates that the first change, the increase, correctly occurs around 2100 cal yr BP (first dashed green density) while the second, the decrease, is thought to occur between 2040-2000 cal yr BP (second dashed green density). While this second changepoint is estimated as occurring slightly more recently than the 2050 cal yr BP in the underlying data generation model, the difference is small. The reasoning behind this minor imprecision can be understood by considering the calibration curve in Fig. 6A. Around 2010 cal yr BP, the calibration curve wiggles upwards, taking a similar value to that at 2050 cal yr BP. This therefore provides another plausible end date and, since there is insufficient evidence of a lack of observed samples in the intervening period, the model is unable to identify the precise changepoint date.

A lack of identifiability is an inherent problem when performing any $^{14}C$-dates-as-data summarisation. Due to the inversions in the calibration curve, there are frequently multiple summarised distributions which could explain the observed data. It is therefore essential to always plot the summary alongside both the $^{14}C$ data and the calibration curve (as in Fig. 6A) to understand if such non-identifiability is a significant issue.

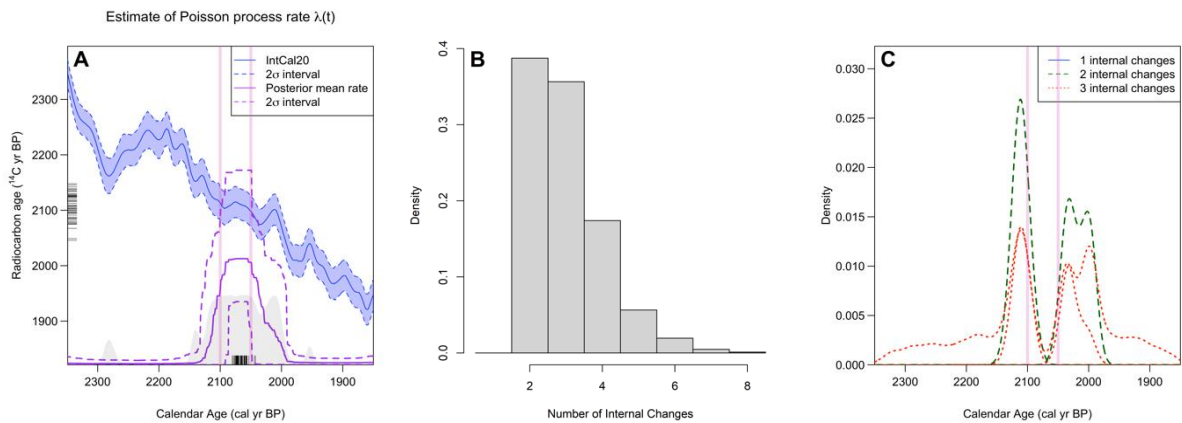

*Figure 6 Results of simulation study 1 where $^{14}C$ samples are simulated from a single uniform phase model. Panel A: Posterior mean of PP occurrence rate $\lambda(\theta)$ with 95% pointwise credible intervals (purple with dashed intervals). The calendar ages of changepoints in the PP used to create the data are shown in shaded light pink. The SPD is shown in light grey. The simulated $^{14}C$ ages are shown as black ticks on the radiocarbon age axis. Panel B: Posterior estimate for the number of internal changes in the rate $\lambda(\theta)$. The underlying model had two. Panel C: Density estimates for the calendar times of the changepoints in $\lambda(\theta)$, conditional on the number of such changes. Underlying changes in model used to create the data are again shown in shaded pink. Note that there are no posterior realisations with just a single changepoint so this is not shown in the plot.*

***Note:*** If one does not know the bounding calendar ages $T_A$ and $T_B$ then these can be estimated from the initial $^{14}C$ observations. It is however important not to choose values that are too distant from the data as the default prior on the changepoint locations prefers changepoints that are spaced somewhat evenly over the entire period. Selecting values of $T_A$ and $T_B$ a long way from the period of interest will therefore mean it is less likely to identify short-term changes (unless the user alters the prior). If not specified by the user, the *carbondate* library default will place $T_A$ and $T_B$ at calendar ages corresponding to the distant tails of the calendar age distributions of the set of $^{14}C$ samples being summarised.

### 5.1.2 Example 2 – A Poisson process with four changepoints

Here we use the dataset first presented in Fig. 5. This was directly simulated from an underlying Poisson process with rate (in events/cal yr):

$$\lambda(\theta) = \begin{cases} h_1 = 0.00 & if\ \theta > 3100\ cal\ yr\ BP, \\ h_2 = 0.08 & if\ 3100 > \theta \geq 2700\ cal\ yr\ BP, \\ h_3 = 0.28 & if\ 2700 > \theta \geq 2300\ cal\ yr\ BP, \\ h_4 = 0.06 & if\ 2300 > \theta \geq 1950\ cal\ yr\ BP, \\ h_5 = 0.00 & if\ \theta < 1950\ cal\ yr\ BP. \end{cases}$$

With this rate $\lambda(\theta)$, we would expect to see 165 events/samples between 3100 and 1950 cal yr BP. Our random simulation creates 171 events, with events more likely to occur in the period from 2700 to 2230 cal yr BP when the rate is highest. We then create corresponding simulated $^{14}$C determinations, $X_1, \ldots, X_{171}$. We apply our PP modelling approach to these determinations assuming bounding calendar ages $T_A = 1750$ cal yr BP and $T_B = 3300$ cal yr BP. The results are presented in Fig. 7.

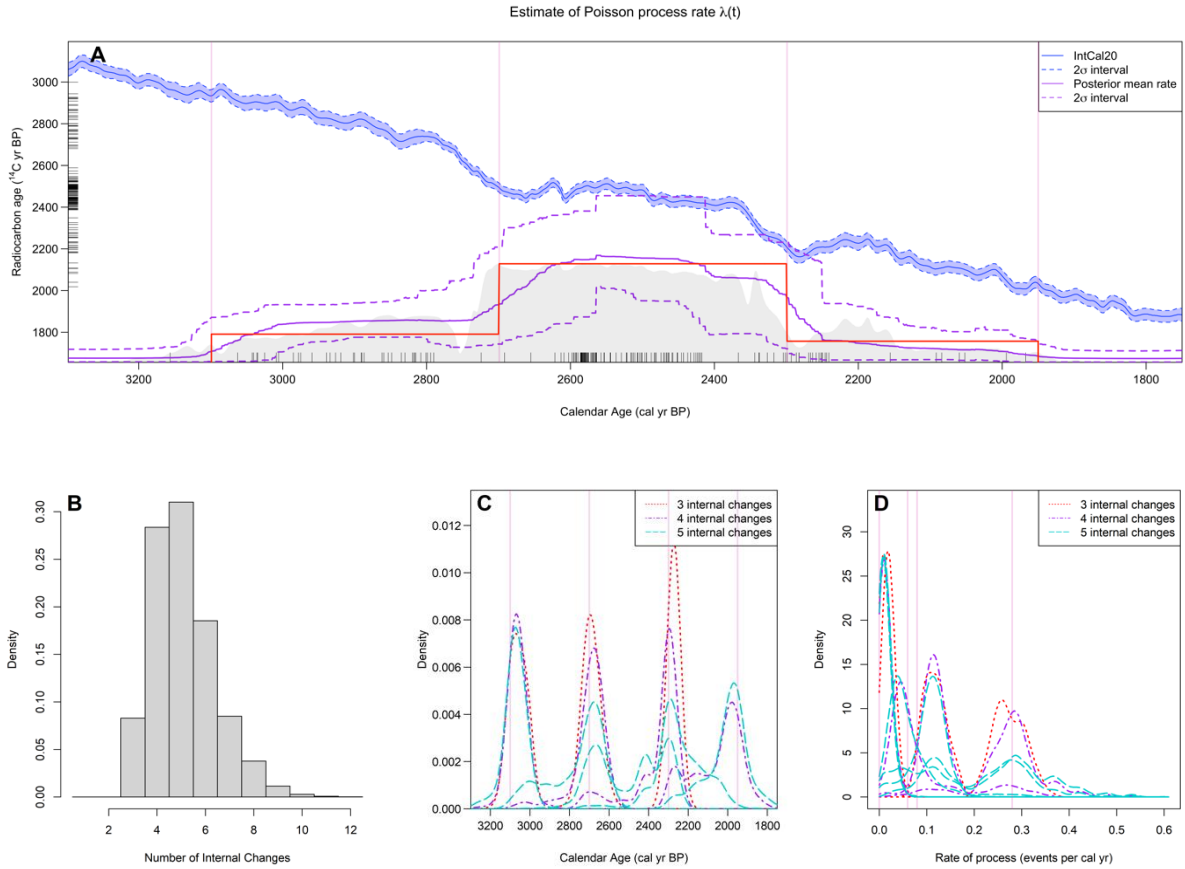

*Figure 7 Results of simulation study 2 where $^{14}$C samples are simulated from a Poisson process model with four changepoints in the occurrence rate. Panel A: Posterior mean of PP occurrence rate $\lambda(\theta)$ with 95% pointwise credible intervals (purple with dashed intervals). The calendar ages of changepoints in the PP used to create the data are shown in shaded light pink. The SPD is shown in light grey. The simulated $^{14}$C ages are shown as black ticks on the radiocarbon age axis. Panel B: Posterior estimate for the number of internal changes in the rate $\lambda(\theta)$. Panel C: Density estimates for the calendar times of the changepoints in $\lambda(\theta)$, conditional on the number of such changes. Panel D: Density estimates for the Poisson process rates in the specific time intervals, conditional on the number of such changes. In panels C and D, the values in the underlying PP model used to simulate the $^{14}$C data are shown in shaded pink.*

We see that our summarisation approach accurately reconstructs the underlying data generation process (Fig. 7A). The posterior mean for $\lambda(\theta)$ demonstrates that we begin with very/few no events occurring, before an initial increase in the occurrence rate of samples occurring between 3100 and 3050 cal yr BP, rising further around 2700 cal yr BP before dropping back down around 2300 cal yr BP and then again back down to near zero just after 2000 cal yr BP. This matches well with the underlying PP model used to generate the $^{14}$C data. Critically, the true PP rate $\lambda(\theta)$ lies entirely within the 95% probability intervals for the posterior mean. In Fig. 7B we see that the RJ-MCMC estimates that the sample occurrence rate exhibits four or five changes (the true model had four). If we condition on three, four, or five changes we see that the RJ-MCMC correctly estimates the times of the changepoints in underlying occurrence rate (Fig. 7C). In those realisations where the MCMC estimates three changepoints, we tend to miss the latest change (which, being a shift from 0.06 to 0 samples per cal yr, is the smallest change). In the case that the MCMC estimates there to be five changepoints, it tends to split either the underlying change in the true $\lambda(\theta)$ at 2700 or 2300 cal yr BP by modelling it as two distinct changepoints located directly adjacent to one another with similar rates in the intervals. This effectively retains the key information that the rate abruptly increases/decreases at these times. We also show in Fig. 7D that the method correctly estimates the specific (piecewise constant) values of the rate $\lambda(\theta)$ in the intervals that it adaptively estimates.

Conversely, the SPD fails to reproduce the underlying model and, without intervals, is very hard to reliably interpret. While it does reflect some general features, it does not recreate key aspects and introduces several spurious features that are not present in the true model. The SPD appears to show a slow increase in samples beginning around 3200 cal yr BP around 100 cal yrs too early. It then erroneously suggests a deep collapse in samples around 2750 cal yr BP – this drop is however simply a consequence of the change in gradient (increased steepness) of the calibration curve rather than the observed $^{14}$C data. The SPD does then correctly increase, remain roughly flat, and then drop around 2300 cal yr BP. However, this drop is rapidly followed by another spurious increase around 2200 cal yr BP that is again an artefact of the method and does not reflect the underlying truth. This comparison between SPD and the fully-Bayesian Poisson process estimate demonstrates the inherent unfitness of SPD approaches to estimate populations from sets of radiocarbon determinations.

### 5.1.3 Example 3 – An Exponential Growth Model

Our final simulation study considers a Poisson process that is equivalent to a model of exponential growth. We consider the specific growth model used in Experiment 1 ($r = 0.03, a = 6000, b = 4000$) of Crema & Shoda (2021). The occurrence rate (in events/cal yr) for the corresponding Poisson process is:

$$\lambda(\theta) = \begin{cases} 0 & if\ \theta > 6000\ cal\ yr\ BP, \\ c \times e^{0.03 \times (6000-\theta)} & if\ 6000 \geq \theta > 4000\ cal\ yr\ BP, \\ 0 & if\ \theta < 4000\ cal\ yr\ BP, \end{cases}$$

where c ($\approx 0.0037$) is chosen so that the Poisson process is expected to generate 500 samples corresponding to the number of samples used by Crema & Shoda (2021). Our particular Poisson process realisation creates 510 calendar age events, each of which is converted to a $^{14}$C determination. We apply our PP modelling approach fixing bounding calendar ages $T_A = 3800$ cal yr BP and $T_B = 6200$ cal yr BP. The results are presented in Fig. 8.

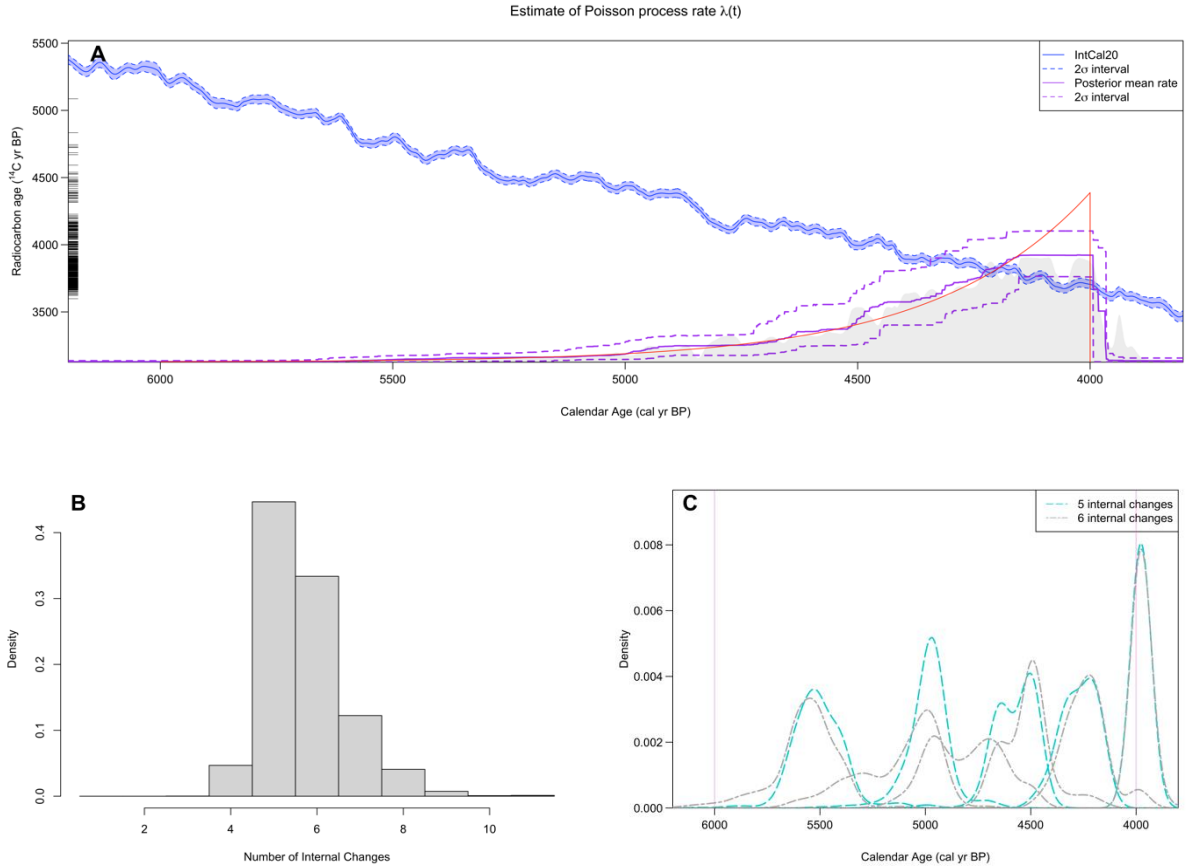

*Figure 8 Results of simulation study 3 where $^{14}C$ samples are simulated from a Poisson process model corresponding to exponential growth. Panel A: Posterior mean of PP occurrence rate $\lambda(\theta)$ with 95% pointwise credible intervals (purple with dashed intervals). The true exponential rate used to generate the samples is shown in red. The SPD is shown in light grey. The simulated $^{14}C$ ages are shown as black ticks on the radiocarbon age axis. Panel B: Posterior estimate for the number of internal changes in the rate $\lambda(\theta)$. Panel C: Density estimates for the calendar times of the changepoints in $\lambda(\theta)$, conditional on the number of such changes. The end points of the underlying exponential growth model values are shown in shaded pink.*

Despite the underlying exponential growth model not having a piecewise constant occurrence rate, our RJ-MCMC method still achieves good reconstruction. The posterior mean occurrence rate increases monotonically over time in accordance with the true rate $\lambda(\theta)$ before correctly dropping to zero at 4000 cal yr BP when the underlying growth model also collapses. Furthermore, the true $\lambda(\theta)$ is almost entirely encapsulated within our 95% probability intervals, the only exception being the very sharp end-peak of growth. This is to be expected since to achieving this extreme peak would require many short rate changes. On the other hand, the SPD again shows multiple peaks and troughs which do not have a clear interpretation as well as indicating a spurious additional peak around 3920 cal yr BP.

The histograms showing the number of changepoints, and the conditional locations of these changes, in our PP posterior are less informative in this instance since the underlying occurrence rate $\lambda(\theta)$ increases continuously. Here, the piecewise constant sections and changes are predominantly used to approximate this exponential increase as indicated by the higher posterior number of changes estimated (compared to the prior mean of 3) and the even spacings of the changepoint locations (which predominantly represent our spaced prior). The collapse at 4000 cal yr BP is however estimated precisely. We do not estimate a changepoint at 6000 cal yr BP since the actual change in $\lambda(\theta)$ here is minimal.

All this is achieved without providing any information to the method about the nature of the underlying generating model. This demonstrates the flexibility of the proposed RJ-MCMC PP approach and its broad applicability.

### 5.2 Real-life Example: Late-Pleistocene Megafauna in N. America

The cause of the collapse of megafauna in N. America during the late-Pleistocene is a fundamental question for ecologists, archaeologists, and anthropologists (Guthrie, 2006). This period of Earth history is characterized by significant climatic changes, it also contains the first known appearance of humans in the region. Specific questions of interest include whether these known environmental changes led to human migration, what is the impact of those same environmental changes on megafauna, and what was the impact of humans on that megafauna? To address these questions, $^{14}$C evidence from Alaska and the Yukon Territory was collected from specimens of varying species (Guthrie, 2006). The $^{14}$C ages for moose (*Alces alces*), bison (*Bison priscus*, which evolved into *Bison bison*), mammoth (*Mammuthus primigenius*), and human occupation are shown in Fig. 1. Potential for collection/sampling biases were considered (see Section 2 for the importance of this to any *dates-as-data* analysis) but the occurrence of the megafauna samples was viewed as being representative of the underlying population sizes since they were accumulated from a very diverse spectrum of circumstances (frequently gold mining operations) and not part of large individual dating projects with specific research foci. The evidence for human occupation consisted of charcoal $^{14}$C dates from discrete identifiable hearths, or food refuse bones, not human skeletons. This should similarly limit potential sampling biases (Guthrie, 2006).

The original analysis only considered the frequency of samples in the radiocarbon age domain and did not consider the essential calibration of these $^{14}$C determinations necessary to place the samples on a calendar scale (Guthrie, 2006). It is not possible to infer changes in the frequency of the samples over calendar time based simply on changes in the density of the $^{14}$C determinations. Variations in the density of $^{14}$C determinations could be a result of changes in the calibration curve's slope rather than in the calendar age frequency of the samples. This is particularly critical for this study as the level of $^{14}$C in the atmosphere undertook considerable changes in the late-Pleistocene (see Fig. 1).

The original $^{14}$C data has been reanalysed incorporating calibration, but only to estimate the extinction times of horse and mammoth (Buck & Bard, 2007). Such an extinction analysis enables only coarse and rudimentary inference. A population may collapse due to an external factor without going entirely extinct. Such detailed, and important, insight into population change is however completely lost if analysis is restricted to a measure of absolute extinction. We wish to obtain greater insight into the interplay between the various species through understanding more nuanced population change. This can be achieved using our PP summarisation approach.

Here, we analyse the occurrence rates of mammoth, bison and moose samples alongside those samples evidencing human occupation. We compare these occurrence rates against one another and the climate, specifically the cold Henrich stadials and Younger-Dryas period (Bard et al., 2013; Waelbroeck et al., 2019). While we do not perform formal modelling of the concurrence (or otherwise) of any changes, several features of the populations which could not previously be rigorously identified are revealed that support theories of subtler human impact and/or ecological replacement or displacement (Guthrie, 2006). We do not show all the varied inference provided by the PP summarisation, but instead highlight key aspects.

For our analysis, we restricted the original $^{14}$C data (Guthrie, 2006) to those samples with $^{14}$C ages between 6000 and 25000 $^{14}$C yr BP to focus on the late-Pleistocene. This left 117 samples relating to mammoth, 64 to bison, 58 to moose, and 46 to human occupation. This relatively small number of samples will mean that we are unable to identify fine-scale variations over time, with the uncertainty on our posterior occurrence rates (quantified as part of our method) remaining larger than if more samples were available. We chose bounding calendar ages $T_A = 6565$ cal yr BP and $T_B = 29430$ cal yr BP to correspond with these $^{14}$C age cut-offs. Our prior on the number of changepoints had an expectation of six changes in rate over this period (little difference was seen when repeating analysis with an expectation of three). The results are shown in Figs. 9 and 10.

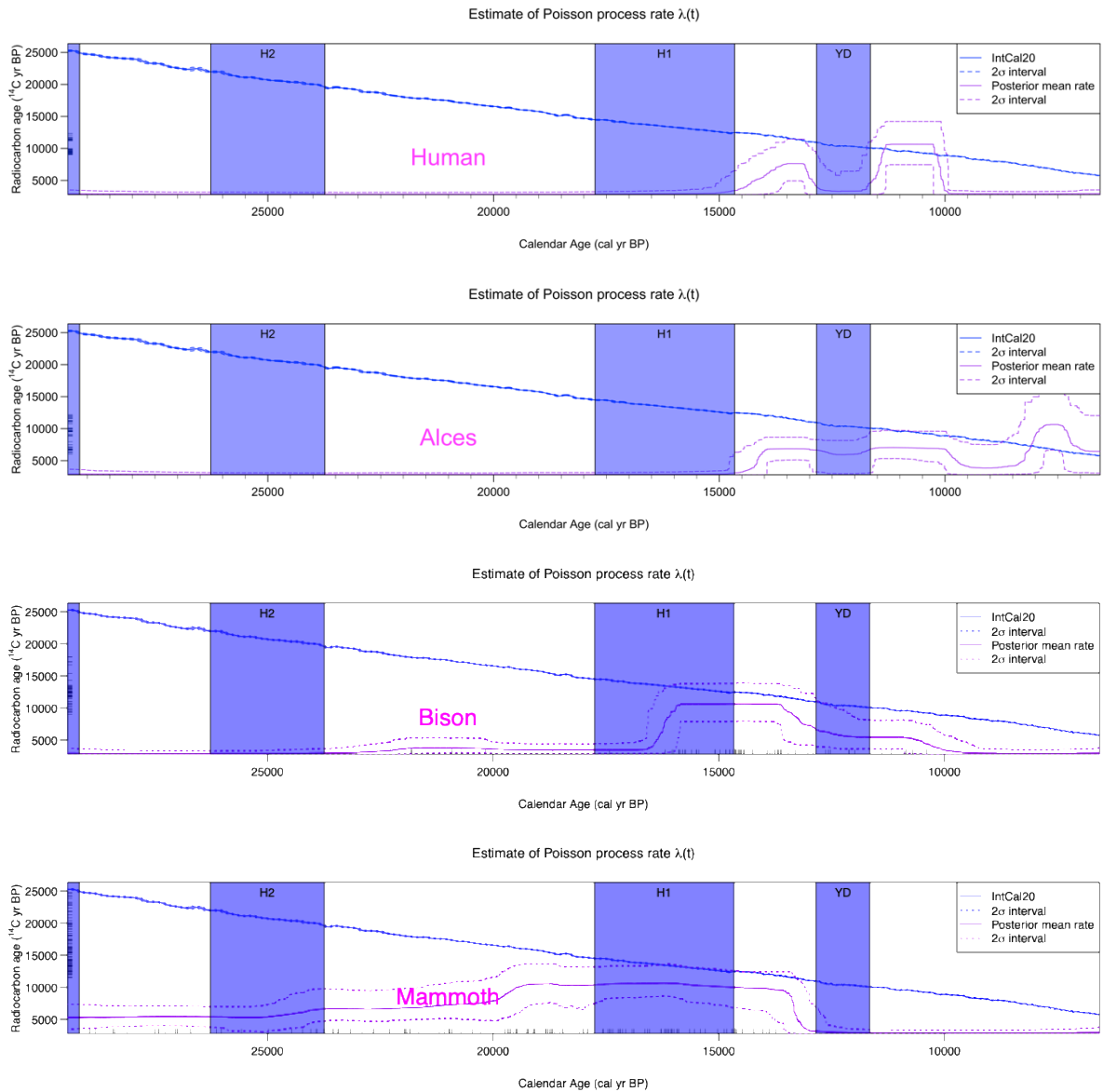

*Figure 9 Occurrence rates for samples showing human occupation alongside alces (moose), bison, and mammoth (Mammuthus primigenius) during the late Pleistocene. Posterior mean of sample occurrence rate is shown in purple (with 95% probability intervals). The $^{14}$C ages are shown on the radiocarbon age axis which were calibrated against the IntCal20 calibration curve (Reimer et al., 2020) shown in blue. The shaded blue time periods correspond to cold stadials – the Younger-Dryas (Y-D) and Heinrich Events 1 and 2 (H1 and H2).*

Humans appeared to arrive in Alaska and the Yukon coincidentally with the end of the cold Heinrich 1 stadial. This is an interesting observation in the context of the peopling of America as, during parts of the cold events, the presence of sea ice may have favoured a coastal route that is probably not sampled in Guthrie's database (Praetorius et al., 2023). Their population/activity increased steadily during this warm period until the onset of the Younger-Dryas (Y-D, another cold stadial) when it rapidly collapsed. Once the Y-D had ended humans rapidly returned to a level almost twice the size of that seen in the previous warm period. Moose appear to have arrived at a similar time to humans. Having arrived, there is little evidence for substantial moose population variation although it too may have dropped during the Younger-Dryas cold period (as evidenced by the widening probability intervals on the posterior occurrence rate).

The population of bison appeared to significantly increase during the middle of the cold Heinrich 1 stadial. Again, this is interesting, although perhaps somewhat counter-intuitive. The Heinrich 1 stadial (18 – 14.5 cal kyr BP) is characterized by some variability, notably two phases, H1a & H1b (Bard et al., 2000) with the youngest phase H1a corresponding to the Heinrich event *sensu stricto* with icebergs surges in the North-Atlantic. The H1 stadial is also complex in the North Pacific (Praetorius et al., 2023). In addition, the population of bison was likely dependent on both the climate conditions (ice and vegetation on land) and the presence of human hunter-gatherers. If humans were migrating via the coastal route during the 16 – 15 cal kyr BP period, the bison may have preferred the mainland of Yukon and Alaska. The population of bison then appears to have stayed steady until the arrival of humans when, at around 13 cal kyr BP, it began to steadily decrease. It did not recover even when humans seemed to have left during the Y-D.

Mammoth were present in the region much earlier – slowly increasing in population over time through the Heinrich 2 and Heinrich 1 events unaffected by climatic changes. However, their population saw a rapid collapse around 13000 cal yr BP coincident with an increase in evidence for human occupation. These synchronicities suggest that both bison and mammoth could have been hunted by humans reducing the size of the populations (in the case of mammoth to extinction).

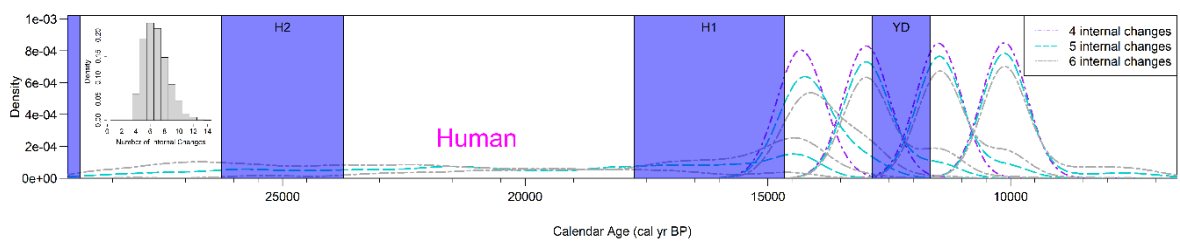

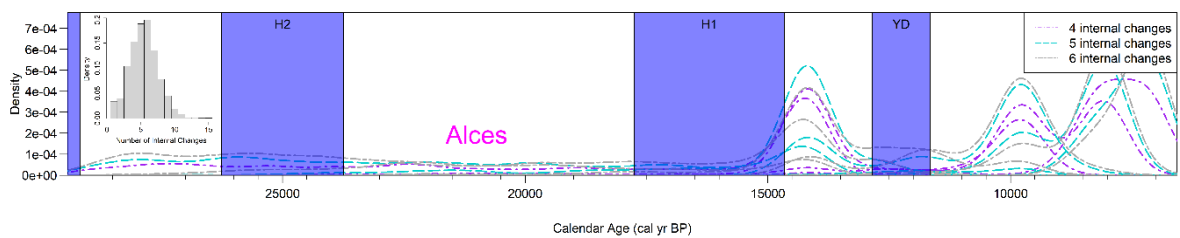
25

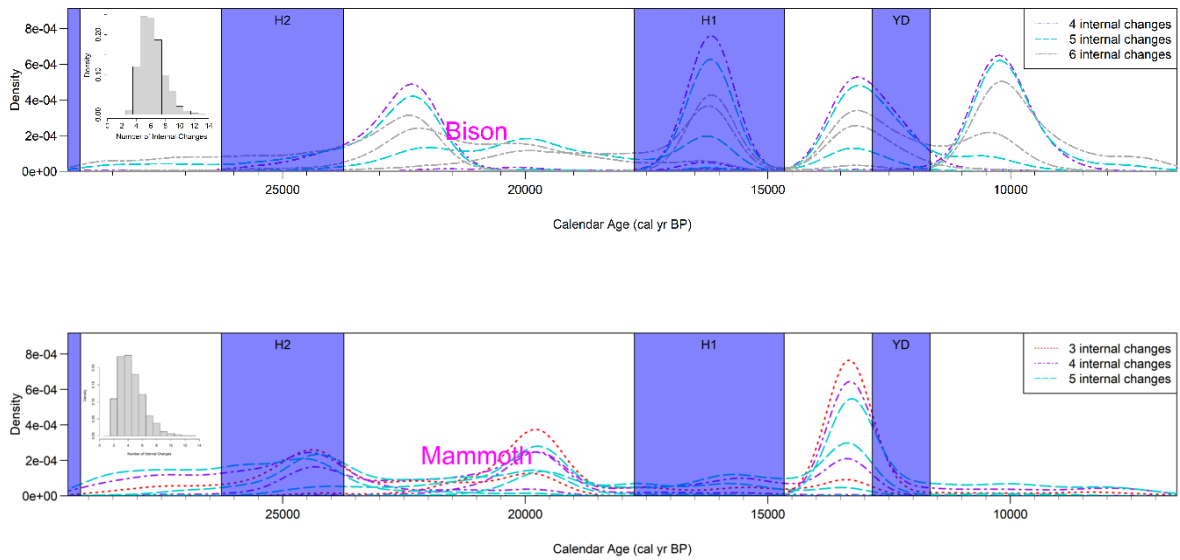

*Figure 10 Estimated number of changepoints in sample occurrence rate (inset histograms) and locations of changepoints (main plots, conditional on their number) for human occupation, alces (moose), bison, and mammoth (Mammuthus primigenius) during the late Pleistocene. For humans, alces and bison we plot the locations conditional on there being 4,5 and 6 changepoints as these are supported by the data (see inset histograms). For mammoth, we plot the locations condition on 3,4 and 5 changepoints (as these have the greatest posterior probabilities).*

# 6 Conclusions

## 6.1 A new Poisson process approach to $^{14}$C summarisation

Big *$^{14}$C-dates-as-data* analyses have become highly popular in the archaeological, environmental, and radiocarbon communities to understand past demographic change (Crema, 2022; Crema & Bevan, 2021; Rick, 1987). However, the most frequently-used approach to perform such analyses, creating summed probability distributions (SPDs), is not rigorous or robust. The inference that such SPDs provide is unreliable and may bear limited relation to the underlying population size/activity that we seek to reconstruct. Attempts to address this through, for example, SPD bootstrap confidence intervals, do not succeed. Indeed, they may make the problem worse by providing spurious confidence in the summary obtained. We therefore advise against the further use of SPDs by the community. While rigorous parametric SPD alternatives, e.g., growth and phase models, have been proposed they limit users to specific forms of summary which must be specified *a priori* (Bronk Ramsey, 2009; Carleton, 2021; Crema & Shoda, 2021; DiNapoli et al., 2021; Price et al., 2021; Timpson et al., 2020). This inherently reduces their flexibility and substantially restricts their field of application. Consequently, these parametric approaches are not able to provide a true SPD replacement.

In this paper we develop a new, rigorous, non-parametric approach to $^{14}$C summarisation that models the occurrence of samples over calendar time as an inhomogeneous Poisson process. We estimate the varying rate of this process (i.e., the underlying sample occurrence rate) and identify specific calendar times where this rate may significantly change. So long as the samples are representative, this sample occurrence rate will provide a proxy for underlying population size/activity. Our novel method addresses questions that archaeologists and environmental scientists have long wanted to answer, allowing them to reliably estimate the calendar age distributions of populations of $^{14}$C samples and critically engage with questions regarding potential underlying structure in these populations.

Our summarisation approach is achieved within a fully-Bayesian RJ-MCMC framework where the sample occurrence rate is estimated jointly, and simultaneously, with calibration of the set of $^{14}$C determinations. Given such a set of samples with $^{14}$C determinations, we obtain estimates for the posterior mean of the sample occurrence rate over calendar time (with rigorous probability intervals) in addition to valuable information on the number of changes in the sample occurrence rate and the calendar times at which such changes occur. This changepoint information enables important further inference if we wish to understand causes for demographic change. All analysis can be performed in R using the *carbondate* library that provides user-friendly plots of key posterior MCMC information (crucially plotting the summary alongside the calibration curve to identify potential issues of non-identifiability). We show in a series of simulated and real-life examples the success of our method.

## 6.2 A DPMM or a PP approach to summarisation

Our PP approach sits alongside the equally-rigorous non-parametric DPMM approach to $^{14}$C summarisation where samples are modelled as belonging to distinct clusters of normal phases within an infinite mixture model (Heaton, 2022). Both methods are available within the *carbondate* library. Intuitively, the approaches are linked even though they approach the problem from different perspectives. The PP implicitly assumes the underlying activity forms a series of uniform phases, while the DPMM assumes a series of normal phases. Which approach, PP or DPMM, is more appropriate will depend upon the dataset under study. If the underlying population is thought to exhibit sharp changes in size of activity or size then the PP approach, with its modelling of the sample occurrence rate as piecewise constant with distinct jumps, may be more suitable. If instead, the population is thought to consist of a smoother evolution where mixtures of groups/cultures slowly rise and fall according, the DPMM may provide more useful inference. We suggest that users might apply both the DPMM and PP approaches and compare outputs as part of the research process.

## 6.3 Future Work

We hope that our work will also raise discussion that big data analysis is challenging, and users must take considerable care – spending time to understand the actual data and its collection. Fundamentally, big data analyses do not remove the critical archaeological requirement that users understand the link between their samples and the question at hand. Particularly key to all *$^{14}$C-dates-as data* analyses, if we wish to use them to understand underlying demographics, is the representativeness of the samples that we are summarising. Work is needed to better understand the sampling biases and taphonomic loss in our existing datasets so that information can be incorporated into analyses (Contreras & Meadows, 2014). Future work could also consider how to best elicit and capture prior, independent, information on the number and location of potential changes in population size/activity.

## Code and Data Availability

All code was implemented in R (R Core Team, 2024) using the *carbondate* library that we have built to accompany the manuscript. The *carbondate* library enables any user to implement the Poisson process summarisation approach on their own datasets. It also enables implementation of the rigorous DPMM summarisation approach (Heaton, 2022) referred to in the text should an approach that more closely mimics SPDs be desired. The library is available through CRAN (https://cloud.r-project.org/web/packages/carbondate/index.html) and on GitHub (https://github.com/TJHeaton/carbondate). A user-guide containing a series of vignettes and worked examples to demonstrate the library's functionality is provided at https://tjheaton.github.io/carbondate/.

The specific scripts to reproduce the simulated and Pleistocene megafauna examples in this paper (which call the *carbondate* library) can be found at https://github.com/TJHeaton/PP-paper-examples.

## Contributions

**TJ Heaton:** Conceptualization, Methodology, Software, Formal Analysis, Writing – Original Draft, Funding acquisition. **S Al-assam**: Methodology, Software. **Bard**: Conceptualization, Writing – Reviewing and Editing.

## Acknowledgements

This work was supported by EPSRC grant EP/X032906/1 and APX\R1\231120. EB is funded by ANR MARCARA. We thank L Wacker, M Christl, P Hoebe, and R McLaughlin for constructive feedback.

# Appendix

## A1 The Poisson Process Model

### A1.1 Inhomogeneous Poisson Process

A Poisson process is often used to model the occurrence of random events. It is described by specifying the random variables $N(s,t)$ for $0 \leq s \leq t$, where $N(s,t)$ represents the number of events that have occurred in the time interval $(s,t]$. Any Poisson process has a parameter $\lambda(\theta)$, known as the rate of the process that controls the expected number of events in a particular time interval. In the case of an inhomogeneous Poisson process, this rate varies over time so that some time periods are expected to have a greater number of event than others.

An (inhomogeneous) Poisson process is then defined by the following two properties:

a) For any $0 \leq s \leq t$, the distribution of $N(s,t)$ is Poisson with parameter $\int_{\theta=s}^{t} \lambda(\theta) \, d\theta$.
b) If $(s_1, t_1], (s_2, t_2], \ldots, (s_n, t_n]$ are disjoint intervals (i.e., non-overlapping) then $N(s_1, t_1), N(s_2, t_2), \ldots, N(s_n, t_n)$ are independent random variables.

In particular, the first of these properties implies that the expected number of events in a specific time interval $(s, t]$ is $\int_{\theta=s}^{t} \lambda(\theta) \, d\theta$. Hence, time periods with higher rates $\lambda(\theta)$ will be expected to have greater number of events. The second implies that, conditional on the rate $\lambda(\theta)$, the number of events that have occurred in one time period is not affected by the number of events in a different, and non-overlapping, time period. It also implies that events/samples are not clustered beyond the effect of the variable rate, e.g., the occurrence of one event does not increase/decrease the probability that another event occurs shortly afterwards. Clustering might occur if a single event of interest generates multiple samples over a short period of time, or makes it more likely there will be another event occurring shortly afterwards.

### A1.2 Modelling the Rate $\lambda(\theta)$ as Piecewise Constant

For our work, we will assume that all the events $\theta_i$ (for $i = 1, \ldots, n$) lie within a section of the timeline $[T_A, T_B]$, and restrict our Poisson process to this time period. Here, $T_A$ cal yr BP represents the lower limit for the calendar age of an event, i.e., the limit for the most recent calendar age; and $T_B$ cal yr BP the upper limit, i.e., the oldest calendar age. In this period, we model the rate $\lambda(\theta)$ of our Poisson process as piecewise constant but with an unknown number of steps each of unknown height. Each step corresponds to a change in the rate $\lambda(\theta)$.

Suppose that there are $k$ such steps, at positions $T_A < s_1 < s_2 < \cdots < s_k < T_B$, and that the rate takes the value $h_j$ for any calendar age $\theta$ between steps $s_j$ and $s_{j+1}$ (i.e., if $s_j \leq \theta < s_{j+1}$). We define $T_A = s_0$ and $T_B = s_{k+1}$ for simplicity). The rate $\lambda(\theta)$ is then specified by the locations of the $k$ steps $\{s_1, s_2, \ldots, s_k\}$ and the heights $\{h_0, h_1, \ldots, h_k\}$ between them.

As is common for reversible jump MCMC (Green, 1995), we assume the true number of steps $k$ (or equivalently the number of changes in the event/sample occurrence rate between $T_A$ and $T_B$) is unknown with a prior drawn from the Poisson distribution,

$$P(Number\ of\ Rate\ Changes = k) = e^{-n_\lambda} \frac{\lambda^k}{k!}$$

conditioned on $k \leq k_{kmax}$. Given $k$, the step positions $s_1, s_2, \ldots, s_k$ are distributed as the even-numbered order statistics from $2k + 1$ points uniformly distributed on $[T_A, T_B]$, and the heights $h_0, h_1, \ldots, h_k$ are independently drawn from the $\Gamma(\alpha, \beta)$ density $\beta^\alpha h^{\alpha-1} e^{-\beta h} / \Gamma(\alpha)$ for $h > 0$. The prior expected number of rate changes is therefore $n_\lambda$, with the prior on the location of these steps aiming to penalise overly-short steps and hence space out the changes in occurrence rate over time.

### A1.3 Modelling the Events and $^{14}$C Determinations

Given our Poisson process and rate $\lambda(\theta)$, we can calculate the likelihood of the events. Firstly, following directly from our definition of a Poisson process, the probability that $n$ events occur between $[T_A, T_B]$ is

$$P(n \; events) = \frac{\left[\int_{T_A}^{T_B} \lambda(\theta) \, d\theta\right]^n}{n!} e^{\int_{T_A}^{T_B} \lambda(\theta) \, d\theta}.$$

Then, conditional on $n$ events having occurred in the period $[T_A, T_B]$, the calendar age $\theta_i$ of each is independently distributed with a density that is proportional to the Poisson rate,

$$f_i(\theta) = \frac{\lambda(\theta)}{\left[\int_{T_A}^{T_B} \lambda(\theta) \, d\theta\right]}.$$

Note that, with our stepwise model on $\lambda(\theta)$, this is equivalent to placing a prior on the calendar age $\theta_i$ of each event that is a combination of multiple uniform phase models. Putting these two components together, the log-likelihood of $\{\theta_i, i = 1, \ldots, n\}$ is

$$l(\lambda | \theta_1, \ldots, \theta_n) = \sum_{i=1}^{n} \log \lambda(\theta_i) - \int_{T_A}^{T_B} \lambda(\theta) \, d\theta$$

We do not however observe the exact calendar ages of each event. Instead, for each, we have a radiocarbon determination,

$$X_i \sim N(\mu(\theta_i), \tau^2(\theta_i) + \sigma_i^2),$$

where $\mu(\cdot)$ and $\tau^2(\cdot)$ are the mean and variance of the IntCal20 calibration curve respectively, and $\sigma_i$ the measurement uncertainty on determination $X_i$.

### A1.4 Final Model

In summary, our approach assumes we have observed $n$ $^{14}$C determinations $\{X_i, i = 1, \ldots, n\}$, each corresponding to an event of interest, which have arisen according to the below statistical model:

$$X_i \sim N(\mu(\theta_i), \sigma_i^2) \; for \; i = 1, \ldots, n$$
$$\theta_1, \ldots, \theta_n \sim PP(\lambda(\theta)) \; with$$
$$\lambda(\theta) = \begin{cases} h_i & if \; s_i \leq \theta < s_{i+1} \\ 0 & otherwise \end{cases}$$
$$k \sim Po(n_\lambda) \quad h_i \sim \Gamma(\alpha, \beta)$$

where $s_i$ represent the $k$ steps $T_a < s_1 < s_2 < \cdots < s_k < T_b$; $\mu(\cdot)$ is the IntCal20 calibration curve; and $\sigma_i$ the measurement uncertainty on determination $X_i$.

We aim to simultaneously estimate both the underlying calendar ages of the events $\theta_1, \ldots, \theta_n$; and the variable rate of the Poisson process $\lambda(\theta)$. For inference on changes in the rate of events over time, it is the rate $\lambda(\theta)$ which is of particular interest.

### A2 The Estimation Process – An MCMC Approach

We use a Metropolis-within-Gibbs approach to estimate the calendar ages $\{\theta_i, i = 1, \ldots, n\}$ and the underlying Poisson process rate $\lambda(\theta)$. Our estimate for $\lambda(\theta)$ will be encapsulated through the locations, and number, of steps $\{s_1, s_2, \ldots, s_k\}$; and the corresponding heights $\{h_0, h_1, \ldots, h_k\}$ of the rate between each step. Our sampler consists of two steps:

1. Update $\theta_i \mid X_i, \lambda(\theta)$ for $i = 1, \ldots, n$
2. Update $\lambda(\theta) \mid \theta_1, \ldots, \theta_n, \alpha, \beta, \lambda$

Updating $\theta_i \mid X_i, \lambda(\theta)$ for $i = 1, \ldots, n$

Given $\lambda(\theta)$, we have an implied prior on $\theta_i$ which is proportional to the rate of the underlying Poisson process at that calendar time:

$$f_i(\theta) = \frac{\lambda(\theta)}{\left[\int_{T_A}^{T_B} \lambda(\theta) \, d\theta\right]}.$$

The posterior for $\theta_i \mid X_i, \lambda(\theta)$ is therefore

$$\pi(\theta_i \mid X_i, \lambda(\theta)) \propto [X_i \mid \theta_i][\theta_i \mid \lambda]$$
$$\propto f(X_i \mid \theta_i) f_i(\theta_i)$$
$$= \varphi(X_i; m(\theta_i), \sigma_i^2 + \tau(\theta_i)^2) \lambda(\theta_i).$$

Here $\varphi(\cdot; A, B^2)$ is the density of a normal distribution with mean $A$ and variance $B^2$; while $m(\theta)$ and $\tau(\theta)^2$ are the published IntCal20 mean and variance at $\theta$ cal yr BP respectively. We can sample directly from this posterior.

Furthermore, the necessary repetition of this update step can be done very quickly if we retain, on a grid, the values of $\varphi(X_i; m(\theta_i), \sigma_i^2 + \tau(\theta_i)^2)$. We need then only reweight these according to our updated $\lambda(\theta)$ each time this update is required.

Updating $\lambda(\theta) \mid \theta_1, \ldots, \theta_n, \alpha, \beta, \lambda$

This update requires reversible jump MCMC. We do not give full details here but refer to the original work of Green (1995). Given the calendar ages, our update here is the precise analogue of the worked example of coal mining disasters (Green, 1995).

In brief, this RJ-MCMC update to our stepwise $\lambda(\theta)$ considers four possible changes to the current rate estimate: a change in the location of a step $s_j$; a change in a particular height $h_j$; the addition of another step change in the rate; or the removal of a step change. The latter two require a change in the dimension of the parameter space (they propose a change to the number of steps and heights in our model for $\lambda(\theta)$) and so the acceptance probabilities in our sampler must be modified accordingly.

## A3   Model Output and Additional Information

### A3.1   Posterior Realisations

Our MCMC provides a set of posterior realisations of $\lambda(\theta)$. Specifically, each posterior realisation provides a value for $k$ (the number of changepoints in the sample occurrence rate) as well as $s_1, s_2, \ldots, s_k$ (the locations of those changes) and $h_1, \ldots, h_{k+1}$ (the rates in each interval). These completely define the sample occurrence rate for that realisation. A set of posterior realisations for the simulated Example 1 – A single uniform phase (A Poisson process with two changepoints) that we described in Section 5.1.1 is shown in Figure S1A. These posterior realisations enable the users to access a large amount of potentially useful information. We expect that the most useful for inference will be the posterior mean of the sample occurrence rate; and information on potential changepoints in that rate via the posterior distribution on the number of changepoints in the sample occurrence rate, and the posterior distribution for the calendar age of those changes.

### A3.2 Posterior Mean Occurrence Rate

To estimate the posterior mean occurrence rate at any calendar time $\theta$, we take the pointwise mean of a large set of these posterior realisations (this set can also be used to provide probability intervals for the posterior mean). Note that, while each individual posterior rate realisation will have completely discontinuous/abrupt changes in rate over calendar time, the posterior mean rate will not. This is because the large number of realisations it averages over will likely all have somewhat different changepoint times (as well as potentially different changepoint numbers). This is a feature of the model whereby, where we are able to average over both the number of changepoints and their location. This provides the method with more flexibility than simply application to data where the occurrence rate is piecewise constant – the underlying discontinuous and piecewise constant model for the occurrence rate can be used as a tool to investigate smoother changes in the rate.

### A3.3 Number and Calendar Timing of Changepoints in Occurrence Rate

Information on the specific changepoint times can be most easily understood by conditioning on the value of $k$ (e.g., restricting to those posterior realisations of $\lambda(\theta)$ that have three changes) for values of $k$ that are supported by the data. We therefore advise users to commence changepoint inference by plotting the histogram of the posterior number of changepoints (e.g. Figure 6B which indicates that two or three changepoints in the rate are the most probable). This should allow users to then select suitable values of $k$ to condition on, e.g., Figure 6C where we show the changepoint times conditional on their being two and three changes. Note that these plots showing the changepoint times become busier, and more challenging to interpret, as the value of $k$ we condition on increases.

### A3.4 Posterior Mean Rate Conditional on Number of Changes

We are also able to plot the posterior mean rate conditional on a user-specified number of changepoints in the rate, i.e., the value of $k$. See Figure S1B. However, we do not recommend this except in very specific circumstances where users know *a priori* the number of changes in rate, but not their timings. Such a situation is unlikely for most applications. We instead recommend that users incorporate the uncertainty in the number of changepoints when calculating the posterior mean occurrence rate as in our default rate plots, e.g., Figure 6A.

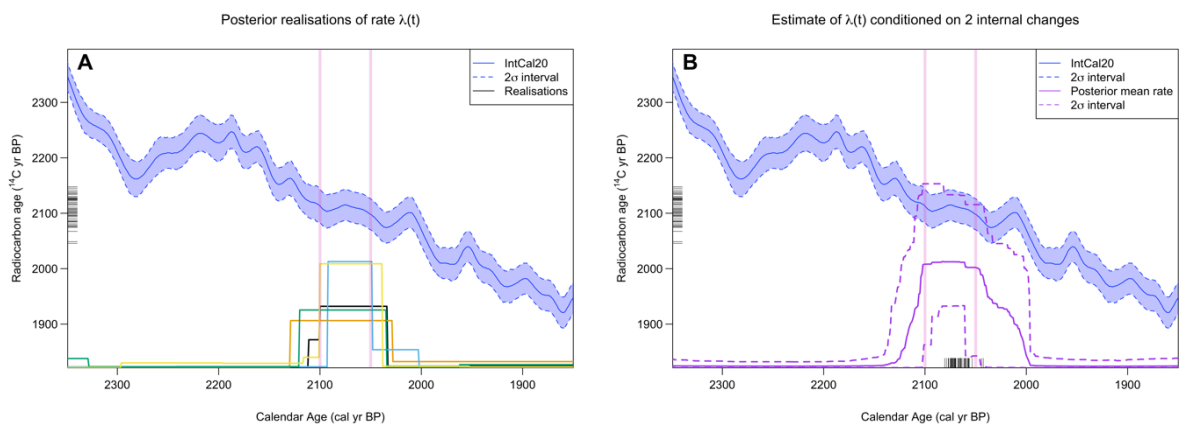

*Figure S1 Panel A: Individual posterior realisations (variously coloured lines) of the sample occurrence rate $\lambda(\theta)$ taken from the simulated Example 1 in Section 5.1.1 (A single uniform phase, or equivalently a Poisson process with two changepoints). Panel B: The posterior mean of the occurrence $\lambda(\theta)$ conditioned on their being two changepoints (i.e., fixing $k = 2$) in the chosen time interval.*

## A4 Selecting model hyperparameters and priors

Within our PP model, we are required to place some priors on the values of parameters. Our default choice of prior means that our approach will prefer to fit parsimonious models where there are fewer changes in the occurrence rate (and changes do not occur in rapid succession) if such models are equally likely to explain the observed $^{14}$C determinations. This should help prevent over-fitting to random sampling variations and reduce the introduction of spurious rate changes. Users are however free to edit these choices if they have independent information.

Prior on the number of changepoints $k \sim Po(n_\lambda)$

The mode allows requires one to specify the prior expected number of changepoints, i.e., $n_\lambda$. Our library default is to select $n_\lambda = 3$. However, this can be easily altered by a user and some checking of robustness/sensitivity to this choice is recommended. Indeed, for the real-life late-Pleistocene megafauna analysis in Section 5.2, we increased this value to $n_\lambda = 6$ as this appeared to better represent the range of values supported by the data (although the final estimates were robust to this choice).

To help select a suitable value in general, we suggest considering the posterior histogram of the number of changepoints and ensuring this has not moved too far from the value chosen for the prior mean. We would also recommend choosing a value for $n_\lambda$ that is somewhat smaller than the maximum number of changes one might be expecting in order not to overfit and identify spurious changepoints.

Prior on the heights of occurrence rates $h_i \sim \Gamma(\alpha, \beta)$

Our default is to select the shape $\alpha = 1$ and the rate $\beta = \frac{T_B - T_A}{n}$ as the hyperparameters for this prior, where $n$ is the total number of samples we are summarising, and $T_B - T_A$ the bounding calendar ages for those samples. This is equivalent to selecting a prior exponential distribution, $h_i \sim Exp(mean = \frac{n}{T_B - T_A})$ where the prior mean value for $h_i$ corresponds to a Poisson process that would expect to generate $n$ samples/events of interest. This choice of exponential prior aims to ensure we do not overfit to the data and instead somewhat shrink the occurrence rates towards zero. As above, it can be altered by the user.